\documentclass[english]{IEEEtran}
\usepackage[T1]{fontenc}
\usepackage[latin9]{inputenc}
\usepackage{amsthm}
\usepackage{amsmath}
\usepackage{amssymb}
\usepackage{esint}
\usepackage{graphicx,epsfig}

\makeatletter
\theoremstyle{plain}
\newtheorem{thm}{\protect\theoremname}
\theoremstyle{remark}
\newtheorem{rem}[thm]{\protect\remarkname}
\theoremstyle{plain}
\newtheorem{cor}[thm]{\protect\corollaryname}
\theoremstyle{plain}
\newtheorem{lem}[thm]{\protect\lemmaname}

\usepackage{cite}
\usepackage{setspace}

\makeatother

\usepackage{babel}
\providecommand{\corollaryname}{Corollary}
\providecommand{\lemmaname}{Lemma}
\providecommand{\remarkname}{Remark}
\providecommand{\theoremname}{Theorem}

\begin{document}

\title{Towards a Simple Relationship to Estimate the Capacity of Static
and Mobile Wireless Networks}

\author{Guoqiang Mao, \emph{Senior Member, IEEE, }Zihuai Lin,\emph{ Senior
Member, IEEE, }Xiaohu Ge\emph{, Senior Member, IEEE, }and Yang Yang\emph{,
Senior Member, IEEE }%
\thanks{G. Mao is with the School of Electrical and Information Engineering,
the University of Sydney and National ICT Australia. %
}\emph{}%
\thanks{Z. Lin is with the School of Electrical and Information Engineering,
the University of Sydney. %
}\emph{}%
\thanks{X. Ge is with the Department of Electronics and Information Engineering,
Huazhong University of Science and Technology. %
}\emph{}%
\thanks{Y. Yang is with Shanghai Research Center for Wireless Communications,
Shanghai Institute of Microsystem and Information Technology,
ShanghaiTech University, Chinese Academy of Sciences. %
}\emph{}%
\thanks{Guoqiang Mao's work is supported by Australian Research Council Discovery
projects DP110100538 and DP120102030. Zihuai Lin's work is supported
by Australian Research Council Discovery Project DP120100405, Linkage
Project LP11010011 and the University of Sydney bridging support grant.
Xiaohu Ge's work is supported by the National Natural Science Foundation
of China (NSFC) under the grants 60872007 and 61271224, NFSC Major
International Joint Research Project under the grant 61210002, the
Ministry of Science and Technology (MOST) of China under the grant
0903, the Hubei Provincial Science and Technology Department under
the grant 2011BFA004. Yang Yang's work is partially supported by the
National Natural Science Foundation of China (NSFC) under the grant
61231009, and by the Science and Technology Commission of Shanghai
Municipality (STCSM) under the grant 11JC1412300.%
}}
\maketitle
\begin{abstract}
Extensive research has been done on studying the capacity of wireless
multi-hop networks. These efforts have led to many sophisticated and
customized analytical studies on the capacity of particular networks.
While most of the analyses are intellectually challenging, they lack
universal properties that can be extended to study the capacity of
a different network. In this paper, we sift through various capacity-impacting
parameters and present a simple relationship that can be used to estimate
the capacity of both static and mobile networks. Specifically, we
show that the network capacity is determined by the average number
of simultaneous transmissions, the link capacity and the average number
of transmissions required to deliver a packet to its destination.
Our result is valid for both finite networks and asymptotically infinite
networks. We then use this result to explain and better understand
the insights of some existing results on the capacity of static networks,
mobile networks and hybrid networks and the multicast capacity. The
capacity analysis using the aforementioned relationship often becomes
simpler. The relationship can be used as a powerful tool to estimate
the capacity of different networks. Our work makes important contributions
towards developing a generic methodology for network capacity analysis
that is applicable to a variety of different scenarios. \end{abstract}
\begin{IEEEkeywords}
Capacity, mobile networks, wireless networks
\end{IEEEkeywords}

\section{Introduction\label{sec:Introduction}}

\IEEEPARstart{W}{ireless} multi-hop networks, in various forms, e.g. wireless sensor
networks, underwater networks, vehicular networks, mesh networks and
unmanned aerial vehicle formations, and under various names, e.g.
ad-hoc networks, hybrid networks, delay tolerant networks and intermittently
connected networks, are being increasingly used in military and civilian
applications. 

Studying the capacity of these networks is an important problem. Since
the seminal work of Gupta and Kumar \cite{Gupta00the}, extensive
research has been done in the area. Particularly, in \cite{Gupta00the}
Gupta and Kumar considered an ad-hoc network with a total of $n$
nodes uniformly and \emph{i.i.d.} on an area of unit size. Furthermore,
each node is capable of transmitting at $W$ bit/s and using a fixed
and identical transmission range. They showed that the transport capacity
and the achievable per-node throughput, when each node randomly and
independently chooses another node in the network as its destination,
are $\Theta\left(W\sqrt{\frac{n}{\log n}}\right)$ and $\Theta\left(\frac{W}{\sqrt{n\log n}}\right)$
respectively%
\footnote{The following notations are used throughout the paper. For two positive
functions $f\left(x\right)$ and $h\left(x\right)$:
\begin{itemize}
\item $f\left(x\right)=o\left(h\left(x\right)\right)$ iff (if and only
if) $\lim_{x\rightarrow\infty}\frac{f\left(x\right)}{h\left(x\right)}=0$;
\item $f\left(x\right)=\omega\left(h\left(x\right)\right)$ iff $h\left(x\right)=o\left(f\left(x\right)\right)$;
\item $f\left(x\right)=\Theta\left(h\left(x\right)\right)$ iff there exist
a sufficiently large $x_{0}$ and two positive constants $c_{1}$
and $c_{2}$ such that for any $x>x_{0}$, $c_{1}h\left(x\right)\geq f\left(x\right)\geq c_{2}h\left(x\right)$;
\item $f\left(x\right)\sim h\left(x\right)$ iff $\lim_{x\rightarrow\infty}\frac{f\left(x\right)}{h\left(x\right)}=1$;
\item $f\left(x\right)=O\left(h\left(x\right)\right)$ iff there exist a
sufficiently large $x_{0}$ and a positive constant $c$ such that
for any $x>x_{0}$, $f\left(x\right)\leq ch\left(x\right)$;
\item $f\left(x\right)=\Omega\left(h\left(x\right)\right)$ iff $h\left(x\right)=O\left(f\left(x\right)\right)$;
\item An event $\xi$ is said to occur almost surely if its probability
equals one;
\item An event $\xi_{x}$ depending on $x$ is said to occur asymptotically
almost surely (a.a.s.) if its probability tends to one as $x\rightarrow\infty$.
\end{itemize}
The above definition applies whether the argument $x$ is continuous
or discrete, e.g. assuming integer values.%
}. When the nodes are optimally and deterministically placed to maximize
throughput, the transport capacity and the achievable per-node throughput
become $\Theta\left(W\sqrt{n}\right)$ and $\Theta\left(\frac{W}{\sqrt{n}}\right)$
respectively. In \cite{Franceschetti07Closing}, Franceschetti \emph{et
al.} considered essentially the same random network as that in \cite{Gupta00the}
except that nodes in the network are allowed to use two different
transmission ranges. The link capacity between a pair of directly
connected nodes is determined by their SINR through the Shannon\textendash{}Hartley
theorem. They showed that by having each source-destination pair transmitting
via the so-called ``highway system'', formed by nodes using the
smaller transmission range, the transport capacity and the per-node
throughput can also reach $\Theta\left(\sqrt{n}\right)$ and $\Theta\left(\frac{1}{\sqrt{n}}\right)$
respectively even when nodes are randomly deployed. The existence
of such highways was established using the percolation theory \cite{Meester96Continuum}.
In \cite{Grossglauser02Mobility} Grossglauser and Tse showed that
in mobile networks, by leveraging on the nodes' mobility, a per-node
throughput of $\Theta\left(1\right)$ can be achieved at the expense
of unbounded delay. Their work \cite{Grossglauser02Mobility} has
sparked huge interest in studying the capacity-delay tradeoffs in
mobile networks assuming various mobility models and the obtained
results often vary greatly with the different mobility models being
considered, see \cite{Dousse04latency,Gamal04Throughput-Delay,Jacquet12On,Kong08Connectivity,Li09Capacity,Neely05Capacity}
and references therein for examples. In \cite{Chen09Order}, Chen
et al. studied the capacity of wireless networks under a different
traffic distribution. In particular, they considered a set of $n$
randomly deployed nodes transmitting to single sink or multiple sinks
where the sinks can be either regularly deployed or randomly deployed.
They showed that with single sink, the transport capacity is given
by $\Theta\left(W\right)$; with $k$ sinks, the transport capacity
is increased to $\Theta\left(kW\right)$ when $k=O(n\log n)$ or $\Theta\left(n\log nW\right)$
when $k=\Omega\left(n\log n\right)$. Furthermore, there is also significant
amount of work studying the impact of infrastructure nodes \cite{Zemlianov05Capacity}
and multiple-access protocols \cite{Chau11Capacity} on the capacity
and the multicast capacity \cite{Li09Multicast}. We refer readers
to \cite{Haenggi09Stochastic} for a comprehensive review of related
work.

The above efforts have led to many sophisticated and customized analytical
studies on the capacity of particular networks. The obtained results
often vary greatly with even a slight change in the scenario being
investigated. While most of the analyses are intellectually challenging,
they lack universal properties that can be extended to study the capacity
of a different network. In this paper, we sift through these capacity-impacting
parameters, e.g. mobility, traffic distribution, spatial node distribution,
the capability of nodes to adjust their transmission power, the presence
of infrastructure nodes, multiple-access protocols and scheduling
algorithms, and present a simple relationship that can be used to
estimate the capacity of wireless multi-hop networks. In addition
to capacity, delay is also an important performance metric that has
been extensively investigated. In this paper we focus on the study
of capacity. We refer readers to \cite{Dousse04latency,Gamal04Throughput-Delay,Jacquet12On,Kong08Connectivity,Li09Capacity}
for relevant work on delay.

The main contribution of this paper is the development of a simple
relationship for estimating the capacity of wireless multi-hop networks
applicable to various different scenarios. The following is a detailed
summary of our contributions:
\begin{itemize}
\item Considering an arbitrary network, we show that the network capacity
is determined by the link capacity, the average number of simultaneous
transmissions, and the average number of transmissions required to
deliver a packet to its destination;
\item We extend the above relationship for arbitrary networks to random
networks; 
\item We apply our new result to determine the asymptotic capacity of several
typical random networks considered in the literature \cite{Gupta00the,Franceschetti07Closing,Grossglauser02Mobility,Neely05Capacity,Zemlianov05Capacity,Li09Multicast,Chau11Capacity}.
The capacity analysis using the aforementioned relationship often
becomes simpler; 
\item Based on the intuitive understanding gained from our result, we point
out limitations of some existing results and suggest further improvements;
\item Furthermore, using our result, the capacity analysis for different
networks can be transformed into the analysis of the three key parameters,
i.e. the link capacity, the average number of simultaneous transmissions,
and the average number of transmissions required to deliver a packet
to its destination. Therefore our work makes important contributions
towards developing a generic methodology for network capacity analysis
that is applicable to a variety of different scenarios. 
\end{itemize}
The rest of the paper is organized as follows: Section \ref{sec:Network-Models}
gives a formal defi{}nition of the network models and notations considered
in the paper. Section \ref{sec:Capacity-of-Static-Networks} gives
the main results in this paper on the capacity of arbitrary networks
and random networks. In Section \ref{sec:Applicability of the result},
we demonstrate wide applications of our result by using it to analyze
the asymptotic capacity of various random networks considered in the
literature \cite{Gupta00the,Franceschetti07Closing,Grossglauser02Mobility,Neely05Capacity,Zemlianov05Capacity,Li09Multicast,Chau11Capacity}.
Finally Section \ref{sec:Conclusion-and-Further} concludes this paper.

\section{Network Models\label{sec:Network-Models}}

We consider two classes of networks in this paper: \emph{arbitrary
networks} and \emph{random networks}.

\subsection{Arbitrary networks}

We use the term arbitrary network to refer to a network with a total
of $n$ nodes arbitrarily and deterministically (i.e. not randomly)
placed in a bounded area $A$ initially. These nodes may be either
stationary or moving following arbitrary and fixed (i.e. not random)
trajectories. A node may choose an arbitrary and fixed number of other
nodes as its destination(s). In the case that a source node has multiple
destination nodes, the source node may transmit the same packets to
its destinations, viz. multicast, or transmit different portions of
its packets to different destinations, viz. unicast. Packets are transmitted
from a source to its destination(s) via multiple intermediate relay
nodes. Each node can be either a source, a relay, a destination or
a mixture. It is assumed that there are always packets waiting at
the source nodes to be transmitted, viz. a so-called saturated traffic
scenario is considered. 

Let $V_{n}$ be the node set. Let $E$ be the set of links. The establishment
of a link between a pair of nodes may follow either the protocol model
or the physical model \cite{Gupta00the}. Our analysis does not depend
on the particular way a link is established. When nodes are mobile,
the link between a pair of nodes may only exist temporarily and the
link set at a particular time instant $t$ may be more appropriately
denoted by $E_{t}$ to emphasize its temporal dependence. In this
paper, we drop the subscript $t$ for convenience. It is assumed that
there is a spatial and temporal path between every source and destination
pair. 

Without loss of generality\cite{Gupta00the,Grossglauser02Mobility,Neely05Capacity,Zemlianov05Capacity,Li09Multicast,Chau11Capacity},
we further assume that each node can transmit at a fixed and known
data rate of $W$ bits per second over a common wireless channel.
Following the same analytical approach as that in \cite{Gupta00the},
it can be shown that it is immaterial to our result if the channel
is broken into several subchannels of capacity $W_{1},W_{2},\cdots,W_{M}$
bits per second, as long as $\sum_{m=1}^{M}W_{m}=W$. This assumption
allows us to ignore some physical layer details and focus on the topological
aspects of the network that determine the capacity. Our result however
can be readily extended to incorporate the situation that each link
has a different and known capacity. We do not consider the impact
of erroneous transmissions in our analysis. Transmission errors will
cause a decrease in the effective link capacity and its impact can
be captured in the parameter $W$, which is assumed to be known.

Denote the above network by $G\left(V_{n},E\right)$ and in this paper,
we study the capacity of $G\left(V_{n},E\right)$. 

In the following paragraphs, we give a formal definition of the capacity
of $G\left(V_{n},E\right)$. Let $v_{i}\in V$ be a source node and
let $b_{i,j}$ be the $j^{th}$ bit transmitted from $v_{i}$ to its
destination. Let $d\left(v_{i},j\right)$ be the destination of $b_{i,j}$.
For unicast transmission, $d\left(v_{i},j\right)$ represents single
destination; for multicast transmission, $d\left(v_{i},j\right)$
represents the set of all destinations of $b_{i,j}$. Let $N_{i,T}^{\chi}$
be the number of bits transmitted by $v_{i}$ \emph{and} which reached,
i.e. successfully received by, their respective destinations during
a time interval $\left[0,T\right]$, with $T$ being an arbitrarily
large number. The superscript $\chi\in\Phi$ denotes the spatial and
temporal scheduling algorithm used in the network and $\Phi$ denotes
the set of all scheduling algorithms. If the same bit is transmitted
from a source to multiple destinations, e.g. in the case of multicast,
it is counted as one bit in the calculation of $N_{i,T}^{\chi}$. 

It is assumed that the network is \emph{stable} $\forall\chi\in\Phi$.
A network is called \emph{stable} if and only if for any fixed $n$,
assuming that each node has an infinite queue, the queue length in
any intermediate relay node storing packets in transit does not grow
towards infinity as $T\rightarrow\infty$, or equivalently the long-term
incoming traffic rate into the network equals the long-term outgoing
traffic rate. It is further assumed that there is no traffic loss
due to queue overflow.

The transport capacity when using the spatial and temporal scheduling
algorithm $\chi$, denoted by $\eta^{\chi}\left(n\right)$, is defined
as:

\begin{equation}
\eta^{\chi}\left(n\right)\triangleq\lim_{T\rightarrow\infty}\frac{\sum_{i=1}^{n}N_{i,T}^{\chi}}{T}\label{eq:definition of capacity using phi}
\end{equation}
and the transport capacity of the network is defined as 
\begin{equation}
\eta\left(n\right)\triangleq\max_{\chi\in\Phi}\eta^{\chi}\left(n\right)\label{eq:definition of capacity}
\end{equation}
Obviously $\eta\left(n\right)\geq\eta^{\chi}\left(n\right),\forall\chi\in\Phi$.

An important special case occurs when the scheduling algorithm divides
the transport capacity equally among all source-destination pairs
asymptotically over time. Denote by $\Phi^{f}\subseteq\Phi$ the set
of \emph{fair} scheduling algorithms that divide the transport capacity
equally among all $m$ source-destination pairs asymptotically over
time. The throughput per source-destination pair is defined as
\begin{equation}
\lambda_{m}\triangleq\max_{\chi\in\Phi^{f}}\frac{\eta^{\chi}\left(n\right)}{m}\label{eq:definition of throughput arbitrary network}
\end{equation}

The above definitions of the transport capacity and throughput capacity
are valid for both finite $n$ and asymptotically infinite $n$.

\subsection{Random networks}

In addition to arbitrary networks, random networks have also been
extensively studied in the literature, particularly the asymptotic
properties of random networks as the number of nodes $n$ approaches
infinity \cite{Gupta00the,Franceschetti07Closing,Grossglauser02Mobility,Neely05Capacity,Zemlianov05Capacity,Li09Multicast,Chau11Capacity}.
By a random network, we mean a network with a total of $n$ nodes
and each node is i.i.d. in a bounded area $A$ initially following
a known distribution. If these nodes are mobile, their trajectories
may also be random and i.i.d. A link between a pair of nodes in a
random network may be established following either the protocol model
or the physical model \cite{Gupta00the}. Denote the above random
network by $G_{n}$ to distinguish it from the arbitrary network considered
in the previous subsection.

Given the randomness involved in the problem statement, the above
definitions of throughput capacity for arbitrary networks need to
be modified to account for ``vanishingly small probabilities'' \cite{Gupta00the}.
Particularly, for asymptotic random networks whose number of nodes
$n$ is sufficiently large, we say that under the spatial and temporal
scheduling algorithm $\chi$, the transport capacity of $G_{n}$ is
$\eta^{\chi}\left(n\right)$ if and only if $\eta^{\chi}\left(n\right)$
is the \emph{maximum} transport capacity that can be achieved \emph{asymptotically
almost surely (a.a.s.)} as\emph{ $n\rightarrow\infty$} under $\chi$.
Given the above modification on $\eta^{\chi}\left(n\right)$, the
transport capacity of an arbitrary network defined in (\ref{eq:definition of capacity})
can still be used for random networks. 

The most extensively studied traffic distribution in random networks
involves each node choosing another node independently as its destination
and the transport capacity being divided equally among all source-destination
pairs. In that case, the total number of source-destination pairs
equals $n$ and the capacity of the network is often studied using
the metric known as the \emph{per-node throughput} or the \emph{throughput
capacity}. Denote by $\Phi^{f}\subseteq\Phi$ the set of \emph{fair
}scheduling algorithms that divide the transport capacity equally
among all $n$ source-destination pairs asymptotically over time.
The per-node throughput (or throughput capacity) is defined as
\begin{equation}
\lambda\left(n\right)\triangleq\max_{\chi\in\Phi^{f}}\frac{\eta^{\chi}\left(n\right)}{n}\label{eq:definition of throughput}
\end{equation}

Intuitively, a scheduling algorithm $\chi$ is fair if it divides
the transport capacity equally among all source-destination pairs
asymptotically over time and also distributes traffic evenly across
$A$ such that there is no traffic hot spot. For nodes uniformly and
i.i.d. on $A$ and each node choosing another node independently as
its destination, which is the scenario studied in Sections \ref{sub:Capacity-of-Random}
and \ref{sec:Applicability of the result}, the technique is well
known to establish the (asymptotic) fairness of a scheduling algorithm
$\chi$, or to construct a (asymptotically) fair scheduling algorithm.
It typically involves partitioning $A$ into a set of equal-size sub-areas,
allocating transmission opportunities equally among all sub-areas
and then demonstrating that using $\chi$, the number of source-destination
pairs crossing each sub-area varies by at most a constant factor.
The conclusion readily follows that the throughput obtainable by each
source-destination pair varies by at most a constant factor and each
source-destination pair has access to throughput of the same order
asymptotically, see \cite{Franceschetti07Closing,Li12Capacity} for
examples. The set of scheduling algorithms analyzed in Section \ref{sec:Applicability of the result}
are known to be fair in the sense that \emph{a.a.s.}, each source-destination
pair can achieve a throughput of the same order.

Note that the above definitions of transport capacity and throughput
capacity for random networks are consistent with those in \cite{Gupta00the,Franceschetti07Closing,Grossglauser02Mobility,Neely05Capacity,Zemlianov05Capacity,Li09Multicast,Chau11Capacity}.
Particularly in \cite{Gupta00the}, a throughput capacity of $\lambda\left(n\right)$
bits per second is called \emph{feasible} if there is a spatial and
temporal scheme for scheduling transmissions such that every node
can send $\lambda\left(n\right)$ bits per second on average to its
chosen destination \cite{Gupta00the}. The throughput capacity of
random networks with $n$ node is of order $\Theta\left(f\left(n\right)\right)$
bits per second if there are deterministic constants $c>0$ and $c'<+\infty$
such that
\[
\lim_{n\rightarrow\infty}\Pr\left(\lambda\left(n\right)=cf\left(n\right)\text{ is feasible}\right)=1
\]
\[
\lim\inf_{n\rightarrow\infty}\Pr\left(\lambda\left(n\right)=c'f\left(n\right)\text{ is feasible}\right)<1
\]

\section{Capacity of Arbitrary and Random Networks\label{sec:Capacity-of-Static-Networks}}

In this section, we analyze the capacity of arbitrary networks and
the capacity of random networks respectively.

\subsection{Capacity of Arbitrary Networks\label{sub:Capacity-of-Arbitrary}}

The following theorem on the capacity of arbitrary networks summarizes
a major result of the paper:
\begin{thm}
\label{thm:Capacity relationship for policy pi}Consider an arbitrary
network $G\left(V_{n},E\right)$. Let $\chi$ be the spatial and temporal
scheduling algorithm used in $G\left(V_{n},E\right)$. Let $k^{\chi}\left(n\right)$
be the average number of transmissions required to deliver a randomly
chosen bit to its destination. Let $Y^{\chi}\left(n\right)$ be the
average number of simultaneous transmissions in $G\left(V_{n},E\right)$,
the transport capacity $\eta^{\chi}\left(n\right)$ satisfies:
\begin{equation}
\eta^{\chi}\left(n\right)=\frac{Y^{\chi}\left(n\right)W}{k^{\chi}\left(n\right)}\label{eq:Capacity of mobile and static networks}
\end{equation}
\end{thm}
\begin{IEEEproof}
Recall from Section \ref{sec:Network-Models} that $v_{i}\in V_{n}$
represents a source node, $b_{i,j}$ represents the $j^{th}$ bit
transmitted from $v_{i}$ to its destination(s), $d\left(v_{i},j\right)$,
and $N_{i,T}^{\chi}$ is the number of bits successfully transmitted
by $v_{i}$ during a time interval $\left[0,T\right]$. 

Let $h_{i,j}^{\chi}$ be the number of transmissions required to deliver
$b_{i,j}$ to its destination (or all destination nodes in $d\left(v_{i},j\right)$
in the case of multicast) when the spatial and temporal scheduling
algorithm $\chi\in\Phi$ is used. Let $Y_{t}^{\chi}\left(n\right)$
be the number of simultaneous transmissions in the network $G\left(V_{n},E\right)$
at time $t$. It follows from the definitions of $k^{\chi}\left(n\right)$
and $Y^{\chi}\left(n\right)$ that

\begin{eqnarray}
k^{\chi}\left(n\right) & = & \lim_{T\rightarrow\infty}\frac{\sum_{i=1}^{n}\sum_{j=1}^{N_{i,T}^{\chi}}h_{i,j}^{\chi}}{\sum_{i=1}^{n}N_{i,T}^{\chi}}\label{eq:definition of average number of transmissions}
\end{eqnarray}
and 
\begin{equation}
Y^{\chi}\left(n\right)=\lim_{T\rightarrow\infty}\frac{\int_{0}^{T}Y_{t}^{\chi}\left(n\right)dt}{T}\label{eq:definition of EY}
\end{equation}

Let $\tau_{i,j,l}$, $1\leq l\leq h_{i,j}$ be the time required to
transmit $b_{i,j}$ in the $l^{th}$ transmission and assume that
the transmitting node is active during the entire $\tau_{i,j,l}$
interval. As each node transmits at the same data rate $W$, $\tau_{i,j,l}=\frac{1}{W}$. 

Given the above definitions, we are now ready to prove the theorem.
\begin{rem}
The technique used in the proof is based on first considering the
\emph{total transmission time}, viz. the amount of traffic transmitted,
measured in bits, multiplied by the time required to transmit each
bit, in the network on the individual node level by aggregating the
transmissions at different nodes, viz. $\sum_{i=1}^{n}\sum_{j=1}^{N_{i,T}^{\chi}}\sum_{l=1}^{h_{i,j}^{\chi}}\tau_{i,j,l}$
shown in the latter equations, and then evaluating the total transmission
time in the network on the network level by considering the number
of simultaneous transmissions in the entire network, viz. $\int_{0}^{T}Y_{t}^{\chi}\left(n\right)dt$
shown in the latter equations. Obviously, the two values must be equal.
On the basis of this observation, the theorem can be established.
\end{rem}
At time $T$, the total transmission time during $\left[0,T\right]$
is given by
\begin{equation}
\sum_{i=1}^{n}\sum_{j=1}^{N_{i,T}^{\chi}}\sum_{l=1}^{h_{i,j}^{\chi}}\tau_{i,j,l}+q_{T}^{\chi}=\frac{1}{W}\sum_{i=1}^{n}\sum_{j=1}^{N_{i,T}^{\chi}}h_{i,j}^{\chi}+q_{T}^{\chi}\label{eq:average transmissions}
\end{equation}
where $\sum_{i=1}^{n}\sum_{j=1}^{N_{i,T}^{\chi}}\sum_{l=1}^{h_{i,j}^{\chi}}\tau_{i,j,l}$
accounts for the transmission time for traffic that has reached its
destination and $q_{T}^{\chi}$ accounts for the transmission time
for traffic still in transit at time $T$. 

Let $p_{max}^{\chi}$ be the maximum length, measured in the number
of hops, of all routes in $G\left(V_{n},E\right)$ under $\chi$,
obviously $p_{max}^{\chi}<n$. Furthermore, since the network is stable,
there exists a positive constant $C_{1}$, independent of $T$, such
that the total amount of traffic in transit is bounded by $C_{1}n$.
Therefore 
\begin{equation}
q_{T}^{\chi}\leq\frac{p_{max}^{\chi}C_{1}n}{W}<\frac{C_{1}n^{2}}{W}\label{eq:inequality on maximum queue length}
\end{equation}

On the other hand, the total transmission time during $\left[0,T\right]$
evaluated on the network level equals $\int_{0}^{T}Y_{t}^{\chi}\left(n\right)dt$.
Obviously
\[
\sum_{i=1}^{n}\sum_{j=1}^{N_{i,T}^{\chi}}\sum_{l=1}^{h_{i,j}^{\chi}}\tau_{i,j,l}+q_{T}^{\chi}=\int_{0}^{T}Y_{t}^{\chi}\left(n\right)dt
\]

When $T$ is sufficiently large and the network is \emph{stable},
using (\ref{eq:inequality on maximum queue length}), the amount of
traffic in transit is negligibly small compared with the amount of
traffic that has already reached its destination. Therefore, the following
relationship can be established:

\begin{equation}
\lim_{T\rightarrow\infty}\frac{\sum_{i=1}^{n}\sum_{j=1}^{N_{i,T}^{\chi}}\sum_{l=1}^{h_{i,j}^{\chi}}\tau_{i,j,l}}{\int_{0}^{T}Y_{t}^{\chi}\left(n\right)dt}=1\label{eq:fundamental relation between capacity and average transmission}
\end{equation}

Noting that $\tau_{i,j,l}=\frac{1}{W}$, Equation (\ref{eq:Capacity of mobile and static networks})
follows readily by combing (\ref{eq:definition of capacity using phi}),
(\ref{eq:definition of average number of transmissions}), (\ref{eq:definition of EY})
and (\ref{eq:fundamental relation between capacity and average transmission}).\end{IEEEproof}
\begin{rem}
Equation (\ref{eq:Capacity of mobile and static networks}) can also
be obtained using Little's formula \cite{Kleinrock75Queueing}. Intuitively,
defining the \emph{system} as consisting of the set of all wireless
channels in $G\left(V_{n},E\right)$, the long-term average effective
arrival rate into the system equals $k^{\chi}\left(n\right)\eta^{\chi}\left(n\right)$,
the long-term average amount of traffic in the system equals $Y^{\chi}\left(n\right)$
and the average time in the system equals $\frac{1}{W}$. Equation
(\ref{eq:Capacity of mobile and static networks}) then readily follows
using Little's formula.
\end{rem}
Equation (\ref{eq:Capacity of mobile and static networks}) is obtained
under a very generic setting and is applicable to networks of any
size. It reveals that the network capacity can be readily determined
by evaluating the average number of simultaneous transmissions $Y^{\chi}\left(n\right)$,
the average number of transmissions required for reaching the destinations
$k^{\chi}\left(n\right)$ and the link capacity $W$. The two parameters
$Y^{\chi}\left(n\right)$ and $k^{\chi}\left(n\right)$ are often
related. For example, in a network where each node transmits using
a fixed transmission range $r\left(n\right)$, reducing $r\left(n\right)$
(while keeping the network connected) will cause increases in both
$Y^{\chi}\left(n\right)$ and $k^{\chi}\left(n\right)$ and the converse.
On the other hand, $Y^{\chi}\left(n\right)$ and $k^{\chi}\left(n\right)$
also have their independent significance, and can be optimized and
studied independently of each other. For example, an optimally designed
routing algorithm can distribute traffic evenly and avoid creating
bottlenecks which helps to significantly increase $Y^{\chi}\left(n\right)$
at the expense of slightly increased $k^{\chi}\left(n\right)$ only,
compared with the shortest-path routing. 

The following corollary is an easy consequence of Theorem \ref{thm:Capacity relationship for policy pi}:
\begin{cor}
\label{cor:capacity of arbitrary networks upper bound}Under the same
setting as that in Theorem \ref{thm:Capacity relationship for policy pi},
\[
\eta\left(n\right)=\max_{\chi\in\Phi}\frac{Y^{\chi}\left(n\right)W}{k^{\chi}\left(n\right)}\leq\frac{\max_{\chi\in\Phi}Y^{\chi}\left(n\right)W}{\min_{\chi\in\Phi}k^{\chi}\left(n\right)}
\]

\end{cor}
Corollary \ref{cor:capacity of arbitrary networks upper bound} allows
the two key parameters that determining the capacity of $G\left(V_{n},E\right)$,
viz. $Y^{\chi}\left(n\right)$ and $k^{\chi}\left(n\right)$ to be
studied separately. Parameter $\max_{\chi\in\Phi}Y^{\chi}\left(n\right)W$
is determined by the maximum number of transmissions that can be accommodated
in the network area. Assuming that each node transmits using a fixed
transmission range $r\left(n\right)$, each transmission will then
``consume'' a disk area of radius at least $\frac{C_{2}r\left(n\right)}{2}$
in the sense that two simultaneous active transmitters must be separated
by an Euclidean distance of at least $C_{2}r\left(n\right)$, where
$C_{2}>1$ is a constant determined by the interference model \cite{Gupta00the}.
The problem of finding the maximum number of simultaneous transmissions,
viz. $\max_{\chi\in\Phi}Y^{\chi}\left(n\right)$, can be converted
into one that finds the maximum number of non-overlapping equal-radius
circles that can be packed into $A$ and then studied as a densest
circle packing problem (see \cite{Yang12Connectivity} for an example).
Parameter $Y^{\chi}\left(n\right)$ can also be studied as the transmission
capacity of networks \cite{Weber10An}. For unicast transmission,
$k^{\chi}\left(n\right)$ becomes the average number of hops between
two randomly chosen source-destination pairs and has been studied
extensively \cite{Mao10Probability}. As will also be shown in Section
\ref{sec:Applicability of the result}, $Y^{\chi}\left(n\right)$
and $k^{\chi}\left(n\right)$ can be optimized separately to maximize
the network capacity.

\subsection{Capacity of Random Networks\label{sub:Capacity-of-Random}}

We now consider the capacity of random networks. Note the connection
between random networks and arbitrary networks that an instance of
a random network forms an arbitrary network. The following result
on the capacity of an arbitrary network can be obtained from Theorem
\ref{thm:Capacity relationship for policy pi}.
\begin{cor}
\label{cor:capacity of random networks}Consider a random network
$G_{n}$. Let $\chi\in\Phi^{f}$ be the spatial and temporal scheduling
algorithm used in $G_{n}$. Let $k^{\chi}\left(n\right)$ be the average
number of transmissions required to deliver a randomly chosen bit
to its destination in an instance of $G_{n}$. Let $Y^{\chi}\left(n\right)$
be the average number of simultaneous transmissions in an instance
of $G_{n}$. Both $k^{\chi}\left(n\right)$ and $Y^{\chi}\left(n\right)$
are random numbers associated with a particular (random) instance
of $G_{n}$. If there exist two positive functions $f\left(n\right)$
and $g\left(n\right)$ such that
\[
\Pr\left(\lim_{n\rightarrow\infty}\frac{k^{\chi}\left(n\right)}{f\left(n\right)}=1\right)=1
\]
and
\[
\Pr\left(\lim_{n\rightarrow\infty}\frac{Y^{\chi}\left(n\right)}{g\left(n\right)}=1\right)=1
\]

the throughput capacity $\lambda^{\chi}\left(n\right)$ satisfies:
\begin{equation}
\Pr\left(\lim_{n\rightarrow\infty}\frac{\lambda^{\chi}\left(n\right)}{\frac{g\left(n\right)W}{nf\left(n\right)}}=1\right)=1\label{eq:capacity of random networks}
\end{equation}
\end{cor}
\begin{IEEEproof}
Using the union bound,
\begin{alignat*}{1}
 & 1-\Pr\left(\frac{\lambda^{\chi}\left(n\right)}{\frac{g\left(n\right)W}{nf\left(n\right)}}=1\right)\\
\leq & \left(1-\Pr\left(\frac{k^{\chi}\left(n\right)}{f\left(n\right)}=1\right)\right)+\left(1-\Pr\left(\frac{Y^{\chi}\left(n\right)}{g\left(n\right)}=1\right)\right)
\end{alignat*}
\begin{eqnarray*}
\\
\end{eqnarray*}
The result in the corollary readily follows from Theorem \ref{thm:Capacity relationship for policy pi}.
\end{IEEEproof}
In reality, such two functions $f\left(n\right)$ and $g\left(n\right)$
required by Corollary \ref{cor:capacity of random networks} do not
necessarily exist or are very difficult to find. Therefore asymptotic
capacity of random networks is more commonly studied by investigating
its upper and lower bounds. The following two corollaries give respectively
an upper and a lower bound on the asymptotic capacity of random networks.
These two corollaries are used in Section \ref{sec:Applicability of the result}
to examine the asymptotic capacity of random networks.
\begin{cor}
\label{cor:an upper bound on per-node throughput}Consider a random
network $G_{n}$. Let $\chi\in\Phi^{f}$ be the spatial and temporal
scheduling algorithm used in $G_{n}$. Let $f\left(n\right)$ and
$g\left(n\right)$ be two positive functions such that
\[
\lim_{n\rightarrow\infty}\Pr\left(\min_{\chi\in\Phi^{f}}k^{\chi}\left(n\right)\geq f\left(n\right)\right)=1
\]
and let $g\left(n\right)$ be a function of $n$ such that
\[
\lim_{n\rightarrow\infty}\Pr\left(\max_{\chi\in\Phi^{f}}Y^{\chi}\left(n\right)\leq g\left(n\right)\right)=1
\]

the throughput capacity of $G_{n}$ satisfies:
\begin{equation}
\lim_{n\rightarrow\infty}\Pr\left(\lambda\left(n\right)\leq\frac{g\left(n\right)W}{nf\left(n\right)}\right)=1\label{eq:capacity of random networks-upper bound}
\end{equation}

\end{cor}

\begin{cor}
\label{cor:lower bound on per-node throughput}Consider a random network
$G_{n}$. Let $\chi\in\Phi^{f}$ be the spatial and temporal scheduling
algorithm used in $G_{n}$. Let $f\left(n\right)$ and $g\left(n\right)$
be two positive functions such that
\[
\lim_{n\rightarrow\infty}\Pr\left(k^{\chi}\left(n\right)\leq f\left(n\right)\right)=1
\]
and
\[
\lim_{n\rightarrow\infty}\Pr\left(Y^{\chi}\left(n\right)\geq g\left(n\right)\right)=1
\]

the throughput capacity of $G_{n}$ satisfies:
\begin{equation}
\lim_{n\rightarrow\infty}\Pr\left(\lambda\left(n\right)\geq\frac{g\left(n\right)W}{nf\left(n\right)}\right)=1,\;\;\forall\chi\in\Phi^{f}\label{eq:capacity of random networks-lower bound}
\end{equation}

\end{cor}
As implied in Corollaries \ref{cor:capacity of arbitrary networks upper bound}
and \ref{cor:an upper bound on per-node throughput}, finding the
throughput capacity upper bound of $G_{n}$ is achieved by analyzing
the upper bound of $Y^{\chi}\left(n\right),\;\forall\chi\in\Phi^{f}$,
viz. $\max_{\chi\in\Phi^{f}}Y^{\chi}\left(n\right)$, and then the
lower bound of $k^{\chi}\left(n\right),\;\forall\chi\in\Phi^{f}$,
viz. $\min_{\chi\in\Phi^{f}}k^{\chi}\left(n\right)$, separately.
An upper bound of $\max_{\chi\in\Phi^{f}}Y^{\chi}\left(n\right)$
can usually be found by analyzing the maximum number of simultaneous
transmissions that can be accommodated in $A$, which is in turn determined
by such parameters like SINR threshold or the transmission range,
independent of $\chi$. A lower bound of $\min_{\chi\in\Phi^{f}}k^{\chi}\left(n\right)$
can often be found by analyzing the average number of hops between
a randomly chosen source-destination pair along the shortest path,
which is mainly determined by the network topology and node distribution,
and is independent of $\chi$. Finding the throughput capacity lower
bound of $G_{n}$ often involves using a constructive technique, i.e.
constructing a particular scheduling algorithm $\chi\in\Phi^{f}$
and analyzing the throughput capacity $\lambda^{\chi}\left(n\right)$
under $\chi$ by analyzing the associated parameters $k^{\chi}\left(n\right)$
and $Y^{\chi}\left(n\right)$.

\section{Applications of the Relationship to Determine the Capacity of Random
Networks\label{sec:Applicability of the result}}

In this section, to demonstrate the usage and applicability of our
results developed in Section \ref{sec:Capacity-of-Static-Networks},
we use these results to re-derive some well-known results in the literature
obtained for different networks and through the use of some intellectually
challenging and customized techniques \cite{Gupta00the,Franceschetti07Closing,Grossglauser02Mobility,Neely05Capacity,Zemlianov05Capacity,Li09Multicast,Chau11Capacity}.
Due to the large amount of existing work in the area, it is not possible
for us to include all of them. Therefore the random networks considered
\cite{Gupta00the,Franceschetti07Closing,Grossglauser02Mobility,Neely05Capacity,Zemlianov05Capacity,Li09Multicast,Chau11Capacity}
are chosen as typical examples only. We show that the use of our result
often lead to simpler analysis. Furthermore, through the intuitive
understanding revealed in our result on the interactions of these
capacity-impacting parameters, we point out limitations in some existing
results and suggest further improvement.

\subsection{Capacity of static ad-hoc networks with uniform transmission capability\label{sub:Capacity-of-static-Kumar}}

In \cite{Gupta00the}, Gupta and Kumar first considered a random network
with $n$ nodes uniformly and \emph{i.i.d.} on a unit square $A$
and each node is capable of transmitting at a fixed rate of $W$ bit/s
using a common channel. Every node chooses its destination randomly
and independently of other nodes and transmits using a fixed and identical
transmission range $r\left(n\right)$. Both the protocol model and
the physical model are considered for modeling the interference. As
shown in \cite{Gupta00the}, results obtained assuming the protocol
model can be readily extended to those assuming the physical model.
Therefore, in this paper, we focus on the protocol model only. 

In the protocol model, a direct transmission from a transmitter $v_{i}$
located at $X_{i}$ to a receiver $v_{j}$ located at $X_{j}$ is
successful if the Euclidean distance between $v_{i}$ and $v_{j}$
is smaller than or equal to $r\left(n\right)$ \emph{and} for every
other node $v_{k}$ simultaneously transmitting over the same channel,
$\left\Vert X_{k}-X_{j}\right\Vert \geq\left(1+\triangle\right)r\left(n\right)$
where the parameter $\triangle>0$ defines a guard zone which prevents
a nearby node from transmitting on the same channel at the same time
and $\left\Vert \bullet\right\Vert $ denotes the Euclidean norm.

Given the above setting, it is straightforward to show that each transmitter
defines a disk with a radius equal to $\frac{1}{2}\triangle r\left(n\right)$
and centered at itself such that for the set of concurrent transmitters,
their respective associated disks do not overlap. Therefore, each
transmitter located in $A$ ``consumes'' a disk of area at least
$\frac{1}{4}\pi\left(\frac{1}{2}\triangle r\left(n\right)\right)^{2}=\frac{\pi}{16}\triangle^{2}r^{2}\left(n\right)$
in $A$ (The worst case happens for a transmitter located at the corners
of $A$ where only one quarter of the disk falls in $A$.). It follows
that
\begin{equation}
\max_{\chi\in\Phi}Y^{\chi}\left(n\right)\leq\frac{1}{\frac{\pi}{16}\triangle^{2}r^{2}\left(n\right)}\label{eq:upper bound on E(Y) Kumar}
\end{equation}

We now establish a lower bound of $\min_{\chi\in\Phi^{f}}k^{\chi}\left(n\right)$.
Let $A_{1}$ be a $\frac{1}{4}\times\frac{1}{4}$ square located at
the lower left corner of $A$ and let $A_{2}$ be a $\frac{1}{4}\times\frac{1}{4}$
square located at the upper right corner of $A$. Using the property
that nodes are uniformly and i.i.d. on $A$, it can be shown that
\emph{a.a.s.} the expected fraction of source-destination pairs with
the source located in $A_{1}$ (or $A_{2}$) and the destination located
in $A_{2}$ (or $A_{1}$) equals $2\times\frac{1}{16}\times\frac{1}{16}=\frac{1}{128}$.
The minimum Euclidean distance between these source-destination pairs
is $\frac{\sqrt{2}}{2}$ and thus the minimum number of hops between
these source-destination pairs is $\frac{\sqrt{2}}{2r\left(n\right)}$.
It is then follows that
\begin{equation}
\lim_{n\rightarrow\infty}\Pr\left(\min_{\chi\in\Phi^{f}}k^{\chi}\left(n\right)\geq\frac{\sqrt{2}}{256}\times\frac{1}{r\left(n\right)}\right)=1\label{eq:lower bound on k(n) Kumar}
\end{equation}

Note that $\Phi^{f}\subseteq\Phi$, the following lemma can be obtained
as an easy consequence of Corollary \ref{cor:an upper bound on per-node throughput},
(\ref{eq:upper bound on E(Y) Kumar}) and (\ref{eq:lower bound on k(n) Kumar}).
\begin{lem}
\label{lem:capacity upper bound Gupta Kumar}In the random network
considered by Gupta and Kumar\cite{Gupta00the} and assuming the protocol
model, the per-node throughput satisfies
\[
\lim_{n\rightarrow\infty}\Pr\left(\lambda\left(n\right)\leq\frac{2048\sqrt{2}}{\pi\triangle^{2}}W\frac{1}{nr\left(n\right)}\right)=1
\]

\end{lem}
In Lemma \ref{lem:capacity upper bound Gupta Kumar}, the upper bound
of $\lambda\left(n\right)$ is expressed as a function of the transmission
range $r\left(n\right)$ and an increase in $r\left(n\right)$ will
reduce the upper bound. As the minimum transmission range required
for the network to be \emph{a.a.s.} connected is well known to be
$r\left(n\right)=\sqrt{\frac{\log n+f\left(n\right)}{\pi n}}$ where
$f\left(n\right)=o\left(\log n\right)$ and $f\left(n\right)\rightarrow\infty$
as $n\rightarrow\infty$ \cite{Gupta98Critical}, the conclusion readily
follows that $\lim_{n\rightarrow\infty}\Pr\left(\lambda\left(n\right)\leq\frac{2048\sqrt{2}}{\triangle^{2}}W\frac{1}{\sqrt{\pi n\log n}}\right)=1$.

We now proceed to obtaining a lower bound of $\lambda\left(n\right)$.
The lower bound is obtained constructively. Specifically, using the
scheduling algorithm $\chi\in\Phi^{f}$ presented in \cite{Gupta00the},
we will analyze the associated $k^{\chi}\left(n\right)$ and $Y^{\chi}\left(n\right)$
and then obtain a lower bound of $\lambda^{\chi}\left(n\right)$.
The lower bound obtained under a particular scheduling algorithm is
of course also a lower bound of $\lambda\left(n\right)$. 

We first recall the scheduling algorithm used in \cite{Gupta00the}.
In \cite{Gupta00the}, the network area $A$ is partitioned into a
set of Voronoi cells such that every Voronoi cell contains a disk
of radius $\rho\left(n\right)=\sqrt{\frac{100\log n}{\pi n}}$ and
is contained in a disk of radius $2\rho\left(n\right)$. Packets are
relayed sequentially from a node in a Voronoi cell to another node
in an adjacent Voronoi cell along the Voronoi cells intersecting the
direct line connecting the source and the destination. Denote the
above scheduling scheme by $\chi$.

The following result on a lower bound of $Y^{\chi}\left(n\right)$
is required for obtaining the lower bound of $\lambda^{\chi}\left(n\right)$:
\begin{lem}
\label{lem:number of simultaneous trans lower bound}In the random
network considered by Gupta and Kumar\cite{Gupta00the} and assuming
the protocol model, there exists a small positive constant $c_{1}$
such that the average number of simultaneous transmissions using $\chi$
satisfies
\[
\lim_{n\rightarrow\infty}\Pr\left(Y^{\chi}\left(n\right)\geq c_{1}\frac{n}{\log n}\right)=1
\]

\end{lem}
Note that each Voronoi cell has an area of at most $\frac{400\log n}{n}$.
Therefore the total number of Voronoi cells in $A$ is at least $\frac{n}{400\log n}$.
The result in Lemma \ref{lem:number of simultaneous trans lower bound}
follows readily from \cite[Lemma 4.4]{Gupta00the}%
\footnote{Strictly speaking, the result in \cite[Lemma 4.4]{Gupta00the} was
derived for nodes on the surface of a sphere. However the result can
be readily modified for a planar area with due consideration to the
boundary effect. Thus we ignore the difference and use the result
directly.%
} which states that \emph{a.a.s.} there exists a positive \emph{constant}
$c_{2}$ such that every $\left(1+c_{2}\right)$ slots, each cell
gets at least one slot in which to transmit. 

In addition to Lemma \ref{lem:number of simultaneous trans lower bound}
, we also need the following lemma that provides an upper bound of
$k^{\chi}\left(n\right)$.
\begin{lem}
\label{lem:average number of hops upper bound Kumar}Under the same
setting as that in Lemma \ref{lem:number of simultaneous trans lower bound},
there exists a positive constant $c_{3}$ such that
\[
\lim_{n\rightarrow\infty}\Pr\left(k^{\chi}\left(n\right)\leq c_{3}\sqrt{\frac{n}{\log n}}\right)=1
\]
\end{lem}
\begin{IEEEproof}
In \cite[Lemma 4.4]{Gupta00the}, it was shown that for every line
connecting an arbitrary source-destination pair, denoted by $L$,
and every Voronoi cell $V\in\Gamma_{n}$ where $\Gamma_{n}$ denotes
the set of Voronoi cells, there exists a positive constant $c_{4}$
such that $\Pr\left(L\text{ intersect }V\right)\leq c_{4}\sqrt{\frac{\log n}{n}}$.
Since each Voronoi cell has an area of at least $\frac{100\log n}{n}$,
the maximum number of Voronoi cells is bounded by $\frac{n}{100\log n}$.
Denoting by $N\left(L\right)$ the expected number of cells intersected
by a randomly chosen source-destination line and using the union bound,
it follows from the above results that $N\left(L\right)\leq\frac{c_{4}}{100}\sqrt{\frac{n}{\log n}}$.
This result, together with the result in \cite[Lemma 4.8]{Gupta00the},
which shows that there exists a sequence $\delta\left(n\right)\rightarrow0$
as $n\rightarrow\infty$ such that $\Pr\left(\text{Every cell }V\in\Gamma_{n}\text{ contains at least one node}\right)\geq1-\delta\left(n\right)$,
allow us to conclude that there exists a positive constant $c_{3}=\frac{c_{4}}{100}$
such that
\[
\lim_{n\rightarrow\infty}\Pr\left(k^{\chi}\left(n\right)\leq c_{3}\sqrt{\frac{n}{\log n}}\right)=1
\]

\end{IEEEproof}
Combing the results in Lemmas \ref{lem:number of simultaneous trans lower bound}
and \ref{lem:average number of hops upper bound Kumar}, and also
using Corollary \ref{cor:lower bound on per-node throughput}, the
following result can be shown:
\begin{lem}
\label{lem:per-node capacity lower bound Gupta and Kumar}In the random
network considered by Gupta and Kumar\cite{Gupta00the} and assuming
the protocol model, there exists a positive constant $c_{5}$ such
that the per-node throughput satisfies
\[
\lim_{n\rightarrow\infty}\Pr\left(\lambda\left(n\right)\geq c_{5}W\sqrt{\frac{1}{n\log n}}\right)=1
\]

\end{lem}
Combing Lemmas \ref{lem:capacity upper bound Gupta Kumar} and \ref{lem:per-node capacity lower bound Gupta and Kumar},
conclusion readily follows that \emph{a.a.s.} $\lambda\left(n\right)=\Theta\left(W\sqrt{\frac{1}{n\log n}}\right)$.

In \cite{Gupta00the}, Gupta and Kumar also investigated the capacity
of arbitrary networks and showed that by placing nodes optimally and
deterministically to maximize the capacity, e.g. on grid points, $\lambda\left(n\right)=\Theta\left(W\sqrt{\frac{1}{n}}\right)$.
Realizing that when nodes are optimally placed, a reduced transmission
range of $r\left(n\right)=\Theta\left(\sqrt{\frac{1}{n}}\right)$
is required for the network to be connected. Following a similar analysis
leading to Lemma \ref{lem:capacity upper bound Gupta Kumar} and using
Theorem \ref{thm:Capacity relationship for policy pi} and (\ref{eq:definition of throughput}),
result readily follows that $\lambda\left(n\right)\leq\frac{2048\sqrt{2}}{\pi\left(1+\triangle\right)^{2}}W\frac{1}{nr\left(n\right)}$
and hence $\lambda\left(n\right)=O\left(W\sqrt{\frac{1}{n}}\right)$.
To obtain a lower bound of $\lambda\left(n\right)$, first it can
be shown that when $r\left(n\right)=\Theta\left(\sqrt{\frac{1}{n}}\right)$,
a scheduling algorithm $\chi$ can be easily constructed such that
$Y^{\chi}\left(n\right)=\Theta\left(\sqrt{\frac{1}{n}}\right)$ and
$k^{\chi}\left(n\right)=\Theta\left(\sqrt{\frac{1}{n}}\right)$ (for
example an algorithm that first routes packets along a horizontal
line to a node on the same vertical height as the destination node
and then routes packets along a vertical line to the destination).
Conclusion then follows that $ $$\lambda^{\chi}\left(n\right)=\Theta\left(W\sqrt{\frac{1}{n}}\right)$
and $\lambda\left(n\right)=\Omega\left(W\sqrt{\frac{1}{n}}\right)$.
Combing the lower and the upper bound, results follows using Theorem
\ref{thm:Capacity relationship for policy pi} that for an arbitrary
network with optimally placed nodes, $\lambda\left(n\right)=\Theta\left(W\sqrt{\frac{1}{n}}\right)$. 

The above results on the throughput capacity of arbitrary networks
and random networks unsurprisingly are consistent with those in \cite{Gupta00the}.
In addition to the above rigorous analysis, we also offer the following
intuitive explanation on the capacity results in \cite{Gupta00the}
using the relationship revealed in Section \ref{sec:Capacity-of-Static-Networks}.
In the network considered by Gupta and Kumar, each node transmits
using a fixed and identical transmission range $r\left(n\right)$.
Therefore each transmission consumes a disk area of radius $\Theta\left(r\left(n\right)\right)$
and $Y\left(n\right)=O\left(\frac{1}{r^{2}\left(n\right)}\right)$.
Here we dropped the superscript $\chi$ when we discuss $k\left(n\right)$
and $Y\left(n\right)$ generally and the result does not depend on
a particular scheduling algorithm being used. Furthermore, a scheduling
algorithm can be readily constructed that distributes the transmissions
evenly across $A$ such that $Y\left(n\right)=\Theta\left(\frac{1}{r^{2}\left(n\right)}\right)$.
Given that the average Euclidean distance between a randomly chosen
pair of source-destination nodes equals a constant, independent of
$n$ \cite{Philip07The}, it can be shown that $k\left(n\right)=\Theta\left(\frac{1}{r\left(n\right)}\right)$.
Thus result follows that the throughput capacity $\lambda\left(n\right)=\Theta\left(\frac{W}{nr\left(n\right)}\right)$,
viz. a smaller transmission range will result in a larger throughput.
The minimum transmission range required for a random network to be
\emph{a.a.s.} connected is known to be $r\left(n\right)=\Theta\left(\sqrt{\frac{\log n}{n}}\right)$
while the minimum transmission range required for a network with optimally
and deterministically deployed nodes is known to be $r\left(n\right)=\Theta\left(\sqrt{\frac{1}{n}}\right)$.
Accordingly, the throughput capacity of random networks and arbitrary
networks with optimally placed nodes are $\Theta\left(\frac{W}{\sqrt{n\log n}}\right)$
and $\Theta\left(\frac{W}{\sqrt{n}}\right)$ respectively. Therefore
the $\frac{1}{\sqrt{\log n}}$ factor is the price in reduction of
network capacity to pay for placing nodes randomly, instead of optimally.

\subsection{Capacity of static networks with non-uniform transmission capability}

In\emph{ }\cite{Franceschetti07Closing}, Franceschetti \emph{et al.
}considered a network with $n$ nodes uniformly and \emph{i.i.d.}
on a square of $\sqrt{n}\times\sqrt{n}$. A node $v_{i}$ can transmit
to another node $v_{j}$ directly at a rate of 
\[
R\left(v_{i},v_{j}\right)=\log\left(1+\frac{Pl\left(X_{i,}X_{j}\right)}{N_{0}+\sum_{k\in\Gamma_{i}}Pl\left(X_{k},X_{j}\right)}\right)
\]
where $\Gamma_{i}$ denotes the set of indices of nodes that are simultaneously
active as $v_{i}$, $l\left(X_{i},X_{j}\right)$ denotes the power
attenuation function and $l\left(X_{i},X_{j}\right)=\min\left\{ 1,e^{-\gamma\left\Vert X_{i}-X_{j}\right\Vert }/\left\Vert X_{i}-X_{j}\right\Vert ^{\alpha}\right\} $
with $\gamma>0$ or $\gamma=0$ and $\alpha>2$, and $N_{0}$ represents
the background noise. It is assumed that all nodes transmit at the
same power level $P$. Each node chooses its destination randomly
and independently of other nodes.
\begin{rem}
Strictly speaking, the results derived in Section \ref{sec:Capacity-of-Static-Networks}
can only be used when the link capacity $W$ is fixed. However it
is straightforward to extend these results to study the capacity of
the network considered in \cite{Franceschetti07Closing} where the
link capacity depends on its SINR and is a variable. More specifically,
given the two functions $g\left(n\right)$ and $f\left(n\right)$
defined in Corollary \ref{cor:an upper bound on per-node throughput},
if a third function $h\left(n\right)$ can be found such that $W=O\left(h\left(n\right)\right)$,
it can be readily shown using Corollary \ref{cor:an upper bound on per-node throughput}
that $\lambda\left(n\right)=O\left(\frac{g\left(n\right)h\left(n\right)}{nf\left(n\right)}\right)$.
Similarly, given the two functions $g\left(n\right)$ and $f\left(n\right)$
defined in Corollary \ref{cor:lower bound on per-node throughput},
if a third function $h\left(n\right)$ can be found such that $W=\Omega\left(h\left(n\right)\right)$,
then $\lambda\left(n\right)=\Omega\left(\frac{g\left(n\right)h\left(n\right)}{nf\left(n\right)}\right)$.
\end{rem}
We first introduce the scheduling algorithm used in \cite{Franceschetti07Closing}.
The network area is partitioned into non-overlapping squares of size
$c^{2}$, called cells hereinafter. These cells are grouped into $l^{2}$
non-overlapping sets of cells where $l=2\left(d+1\right)$ and within
each set, adjacent cells are separated by an Euclidean distance of
$\left(l-1\right)c$, see Fig. 4 of \cite{Franceschetti07Closing}
for an illustration. Parameter $d$ is a positive integer to be specified
later. The time is also divided into $l^{2}$ time slots, which are
equally distributed among the $l^{2}$ sets of cells. Within each
time slot, at most one node in a cell can transmit. Furthermore, nodes
located in cells belonging to the same set can transmit at the same
time and nodes located in cells of different sets should use different
time slots to transmit. The following result was established in \cite{Franceschetti07Closing}
on the transmission rate between a pair of directly connected transmitter
and receiver, which will be used in the later analysis:
\begin{lem}
\label{lem:Franceschetti Results on Single Hop}Using the above scheduling
algorithm, for any integer $d>0$, there exists an $W\left(d\right)>0$
such that a.a.s., when a node is scheduled to transmit, the node can
transmit directly to any other node located within an Euclidean distance
of $\sqrt{2}c\left(d+1\right)$ at rate $W\left(d\right)$. Furthermore,
as $d$ tends to infinity, we have 
\[
W\left(d\right)=\Omega\left(d^{-\alpha}e^{-\gamma\sqrt{2}cd}\right)
\]

\end{lem}
Lemma \ref{lem:Franceschetti Results on Single Hop} is essentially
the same as Theorem 3 in \cite{Franceschetti07Closing} except that
in \cite[Theorem 3]{Franceschetti07Closing}, it was considered that
$W\left(d\right)$ is further multiplied by the fraction of time a
cell is scheduled to be active, i.e. $1/l^{2}$, and the data rate
is given in terms of rate per cell whereas in Lemma \ref{lem:Franceschetti Results on Single Hop},
$W\left(d\right)$ corresponds to the link rate, i.e. $W$ in Theorem
\ref{thm:Capacity relationship for policy pi} and Corollaries \ref{cor:capacity of arbitrary networks upper bound},
\ref{cor:capacity of random networks}, \ref{cor:an upper bound on per-node throughput}
and \ref{cor:lower bound on per-node throughput}.

In addition to the above result, capacity analysis in \cite{Franceschetti07Closing}
also relies on the use of the percolation theory. More specifically,
the $\sqrt{n}\times\sqrt{n}$ square is partitioned into $L=\left\lceil \frac{\sqrt{n}}{\kappa\log\left(\sqrt{n}\right)}\right\rceil $
non-overlapping horizontal slabs where $\kappa$ is a positive constant
$ $and each slab is of size $\frac{\sqrt{n}}{L}\times\sqrt{n}$.
By symmetry, the $\sqrt{n}\times\sqrt{n}$ square can also be partitioned
into $L=\left\lceil \frac{\sqrt{n}}{\kappa\log\left(\sqrt{n}\right)}\right\rceil $
non-overlapping vertical slabs and each slab is of size $\sqrt{n}\times\frac{\sqrt{n}}{L}$.
Using the percolation theory, it was shown that there exists positive
constants $c_{1}$ and $c_{2}$ such that by directly connecting nodes
separated by an Euclidean distance of at most $c_{1}$ only, \emph{a.a.s.}
there are at least $c_{2}\log\left(\sqrt{n}\right)$ \emph{disjoint}
left-to-right (top-to-bottom) crossing paths within every horizontal
(vertical) slab as $n\rightarrow\infty$ \cite[Theorem 5]{Franceschetti07Closing}.
These crossing paths are termed ``highway'' in \cite{Franceschetti07Closing}.
Furthermore it was shown that for nodes not part of the highway, they
can access their respective nearest highway node in single hops of
length at most proportional to $\log\left(\sqrt{n}\right)$, i.e.
the Euclidean distance between non-highway nodes and their respective
nearest highway nodes is $O\left(\log\left(\sqrt{n}\right)\right)$.

On the basis of the above results, the following scheduling algorithm
was used in \cite{Franceschetti07Closing} to deliver a packet from
its source to its destination. The algorithm uses four separate phases,
and in each phase time is divided into $l^{2}=4\left(d+1\right)^{2}$
slots where the value of $d$ varies in each phase. The first phase
is used by source nodes to access their nearest highway nodes; in
the second phase, information is transported on the horizontal highways;
in the third phase information is transported on vertical highways
to highway nodes nearest their respective destinations; and in the
fourth phase information is delivered to the respective destinations.
The first and fourth phases use direct transmissions to deliver information
from the source nodes to the respective highway nodes within Euclidean
distance $O\left(\log\left(\sqrt{n}\right)\right)$ away; while the
second and third phases use multiple hops to deliver information hop-by-hop
along the highway and each hop is separated by a maximum Euclidean
distance of $c_{1}$. Denote the above scheduling algorithm by $\xi$.
The following result on the throughput capacity can be established:
\begin{lem}
\label{lem:capacity of ad hoc networks using highway}Using the scheduling
algorithm $\xi$, the throughput capacity in the random networks considered
in \cite{Franceschetti07Closing} satisfies $\lambda^{\xi}\left(n\right)=\Omega\left(\frac{1}{\sqrt{n}}\right)$.\end{lem}
\begin{IEEEproof}
Denote the per-node throughput in the four different phases by $\lambda_{1}^{\xi}\left(n\right)$
, $\lambda_{2}^{\xi}\left(n\right)$, $\lambda_{3}^{\xi}\left(n\right)$
and $\lambda_{4}^{\xi}\left(n\right)$ respectively. We analyze $\lambda_{1}^{\xi}\left(n\right)$
, $\lambda_{2}^{\xi}\left(n\right)$, $\lambda_{3}^{\xi}\left(n\right)$
and $\lambda_{4}^{\xi}\left(n\right)$ separately in the following
paragraphs to obtain $\lambda^{\xi}\left(n\right)$ where $\lambda^{\xi}\left(n\right)=\min\left\{ \lambda_{1}^{\xi}\left(n\right),\lambda_{2}^{\xi}\left(n\right),\lambda_{3}^{\xi}\left(n\right),\lambda_{4}^{\xi}\left(n\right)\right\} $.

We first analyze the link capacity in phase 1. From the earlier result
that the Euclidean distance between non-highway nodes and their respective
nearest highway nodes is $O\left(\log\left(\sqrt{n}\right)\right)$,
there exists a positive constant $c_{3}$ such that \emph{a.a.s.}
the Euclidean distance between non-highway nodes and their respective
nearest highway nodes is smaller than or equal to $c_{3}\log n$.
Choosing the value of $d$ such that $d$ is the smallest integer
satisfying $\sqrt{2}c\left(d+1\right)\geq c_{3}\log n$ and using
Lemma \ref{lem:Franceschetti Results on Single Hop}, it follows that
each non-highway node can transmit to its nearest highway node at
a rate of $\Omega\left(d^{-\alpha}e^{-\gamma\sqrt{2}cd}\right)=\Omega\left(\left(\log n\right)^{-\alpha}n^{-c\gamma\frac{\sqrt{2}}{2}}\right)$
\emph{a.a.s.} using $\xi$. 

Now we analyze the number of simultaneous transmissions in phase 1.
Note that each highway node is separated from its nearest highway
node by at most an Euclidean distance $c_{1}$. Therefore if a node
has no other node located within an Euclidean distance of $c_{1}$
from itself, that node must be a non-highway node. Let $N_{h}$ be
the number of cells where each cell has at least one non-highway node,
let $N_{o}$ be the number of cells where each cell has exactly one
non-highway node and let $N_{iso}$ be the number of cells where each
cell has exactly one node \emph{and} that node has no other node located
within an Euclidean distance of $c_{1}$ from itself. It follows from
the above observation that 
\begin{equation}
N_{h}\geq N_{o}\geq N_{iso}\label{eq:inequality on N_h}
\end{equation}

Now we further analyze the asymptotic property of $N_{iso}$. Let
$\Gamma$ denote the set of all cells. Let $I_{i}$ be an indicator
random variable such that if the $i^{th}$ cell, denoted by $C_{i}$,
has exactly one node \emph{and} that node has no other node located
within an Euclidean distance of $c_{1}$ from itself, $I_{i}=1$;
otherwise $I_{i}=0$. It follows from the definition of $N_{iso}$
that $N_{iso}=\sum_{C_{i}\in\Gamma}I_{i}$. Using the property that
nodes are uniform and i.i.d., it can be shown that $\lim_{n\rightarrow\infty}E\left(I_{i}\right)=p=c^{2}e^{-c^{2}}e^{-\pi c_{1}^{2}}$
where $c^{2}e^{-c^{2}}$ is the probability that $C_{i}$ has exactly
one node and $e^{-\pi c_{1}^{2}}$ is the probability that the node
has no other node located within an Euclidean distance of $c_{1}$
from itself. Furthermore $Var\left(I_{i}\right)=E\left(I_{i}^{2}\right)-E^{2}\left(I_{i}\right)=E\left(I_{i}\right)-E^{2}\left(I_{i}\right)$
and $\lim_{n\rightarrow\infty}Var\left(I_{i}\right)=p-p^{2}$. Note
that $I_{i}$ and $I_{j}$ are asymptotically independent as $n\rightarrow\infty$
if the associated cells $C_{i}$ and $C_{j}$ are separated by an
Euclidean distance greater than or equal to $2c_{1}$. Denote by $\Gamma_{ind}$
a maximal set of cells where adjacent cells are separated by an Euclidean
distance $\mu=\left\lceil \frac{2c_{1}}{c}\right\rceil c$. It can
be readily shown that $\left|\Gamma_{ind}\right|\geq\left(\frac{\sqrt{n}}{\mu+c}\right)^{2}$,
where $\left|\Gamma_{ind}\right|$ denotes the cardinality of $\Gamma_{ind}$.
Therefore using the central limit theorem,

\[
\lim_{n\rightarrow\infty}\Pr\left(\sum_{C_{i}\in\Gamma_{ind}}I_{i}\geq\frac{n}{\left(\mu+c\right)^{2}}-h\left(n\right)\right)=1
\]
where $h\left(n\right)$ is an arbitrary positive function satisfying
$h\left(n\right)=o\left(n\right)$ and $\lim_{n\rightarrow\infty}h\left(n\right)=\infty$.
Noting that $N_{iso}=\sum_{C_{i}\in\Gamma}I_{i}\geq\sum_{C_{i}\in\Gamma_{ind}}I_{i}$
and using inequality (\ref{eq:inequality on N_h}) and the above equation,
\emph{a.a.s. $N_{h}=\Omega\left(n\right)$ }as \emph{$n\rightarrow\infty$.
}Using $\xi$, every $l^{2}=4\left(d+1\right)^{2}$ time slots, each
cell gets one time slot to transmit. Therefore \emph{a.a.s.} the average
number of simultaneous transmissions in phase 1 equals $\Omega\left(\frac{n}{4\left(d+1\right)^{2}}\right)$.

Note that in phase 1, only direct transmission is allowed. It then
follows from Corollary \ref{cor:lower bound on per-node throughput}
that in the first phase, each node can have access a per-node throughput
of $\lambda_{1}^{\xi}\left(n\right)$ where 
\[
\lambda_{1}^{\xi}\left(n\right)=\Omega\left(\frac{n}{4\left(d+1\right)^{2}}\right)\times\frac{\Omega\left(\left(\log n\right)^{-\alpha}n^{-c\gamma\frac{\sqrt{2}}{2}}\right)}{n}
\]
or equivalently $\lambda_{1}^{\xi}\left(n\right)=\Omega\left(\left(\log n\right)^{-\alpha-2}n^{-c\gamma\frac{\sqrt{2}}{2}}\right)$. 

Using a similar analysis, it can be shown that $\lambda_{4}^{\xi}\left(n\right)=\Omega\left(\left(\log n\right)^{-\alpha-2}n^{-c\gamma\frac{\sqrt{2}}{2}}\right)$.

Now we analyze the throughput capacity in phases 2 and 3. We consider
phase 2 first. In phase 2, $d$ is chosen such that $d$ is the smallest
integer satisfying $\sqrt{2}c\left(d+1\right)\geq c_{1}$. It follows
from Lemma \ref{lem:Franceschetti Results on Single Hop}, \emph{a.a.s.}
there exists a positive constant $c_{4}$ such that each highway node
can transmit at a rate of at least $ $$c_{4}$ bits per second, i.e.
$W>c_{4}$ in phase 2.

As introduced earlier, \emph{a.a.s.} each horizontal slab of size
$\frac{\sqrt{n}}{L}\times\sqrt{n}$ has at least $c_{2}\log\left(\sqrt{n}\right)$
\emph{disjoint} highways where $L=\left\lceil \frac{\sqrt{n}}{\kappa\log\left(\sqrt{n}\right)}\right\rceil $.
Two nodes belonging to two disjoint highways are separated by an Euclidean
distance of at least $c_{1}$. Therefore the number of disjoint highways
that can cross a cell is at most $\left\lceil \frac{c^{2}}{\frac{1}{4}\pi c_{1}^{2}}\right\rceil $.
Each horizontal slab has $\frac{\sqrt{n}}{L}\times\frac{\sqrt{n}}{c^{2}}$
cells. Thus each horizontal highway crosses at most $\frac{\sqrt{n}}{L}\times\frac{\sqrt{n}}{c^{2}}\times\left\lceil \frac{c^{2}}{\frac{1}{4}\pi c_{1}^{2}}\right\rceil /\left(c_{2}\log\left(\sqrt{n}\right)\right)=O\left(\sqrt{n}\right)$
cells. A packet moves by at least one cell in each hop. Therefore
the average number of hops traversed by a packet in phase 2 is $O\left(\sqrt{n}\right)$. 

Furthermore, \emph{a.a.s.} the total number of disjoint horizontal
highways is at least $c_{2}L\log\left(\sqrt{n}\right)>\frac{c_{2}}{\kappa}\sqrt{n}$
and each horizontal highway crosses at least $\frac{\sqrt{n}}{c}$
cells where $\sqrt{n}$ is the minimum length of a left-to-right line
in $A$. The number of disjoint highways that can cross a cell is
at most $\left\lceil \frac{c^{2}}{\frac{1}{4}\pi c_{1}^{2}}\right\rceil $.
Therefore, \emph{a.a.s.} the number of cells where each cell contains
at least one high-way node is at least $\frac{c_{2}}{\kappa}\sqrt{n}\times\frac{\sqrt{n}}{c}/\left\lceil \frac{c^{2}}{\frac{1}{4}\pi c_{1}^{2}}\right\rceil $.
Using $\xi$, every $l^{2}=4\left(d+1\right)^{2}$ time slots, each
cell gets one time slot to transmit. It follows that \emph{a.a.s.
}the average number of simultaneous transmissions in phase 2 is greater
than or equal to $\frac{c_{2}}{\kappa}\sqrt{n}\times\frac{\sqrt{n}}{c}\times\frac{1}{l^{2}}/\left\lceil \frac{c^{2}}{\frac{1}{4}\pi c_{1}^{2}}\right\rceil =c_{5}n$,
where $c_{5}=\frac{c_{2}}{\kappa}\times\frac{1}{c}\times\frac{1}{l^{2}}/\left\lceil \frac{c^{2}}{\frac{1}{4}\pi c_{1}^{2}}\right\rceil $
is a positive constant independent of $n$.

It follows from the above analysis and Corollary \ref{cor:lower bound on per-node throughput}
that
\[
\lambda_{2}^{\xi}\left(n\right)=\Omega\left(\frac{1}{\sqrt{n}}\right)
\]

By symmetry, $\lambda_{3}^{\xi}\left(n\right)=\Omega\left(\frac{1}{\sqrt{n}}\right)$
. By choosing the value of $c$ such that $c\gamma\frac{\sqrt{2}}{2}<\frac{1}{2}$,
the conclusion in the lemma readily follows.
\end{IEEEproof}
Lemma \ref{lem:capacity of ad hoc networks using highway} allows
us to conclude that the throughput capacity in the random network
considered by Franceschetti et al. satisfies $\lambda\left(n\right)=\Omega\left(\frac{1}{\sqrt{n}}\right)$,
which is consistent with the result in \cite{Franceschetti07Closing}.

In \cite{Franceschetti07Closing}, essentially nodes are allowed to
use two transmission ranges, viz. a smaller transmission range of
$\Theta\left(1\right)$ for nodes forming the highways and a larger
transmission range of $O\left(\log\left(\sqrt{n}\right)\right)$ for
non-highway nodes to access their respective nearest highway nodes.
Most transmissions are through the highway using the smaller transmission
range while the larger transmission range is only used for the last
mile in phases 1 and 4. It can be shown that phases 1 and 4 do not
become the bottleneck in determining the throughput capacity. Therefore
both $Y\left(n\right)$ and $k\left(n\right)$ are dominated by the
smaller transmission range and accordingly $Y\left(n\right)=\Theta\left(n\right)$,
$k\left(n\right)=\Theta\left(\sqrt{n}\right)$. Furthermore, as a
consequence of Lemma \ref{lem:Franceschetti Results on Single Hop},
$W=\Omega\left(1\right)$. It then readily follows that $\lambda\left(n\right)=\Omega\left(\frac{1}{\sqrt{n}}\right)$.
This higher throughput capacity, compared with that in \cite{Gupta00the},
is achieved by allowing nodes to adjust their transmission capabilities
as required.

In \cite{Chau11Capacity}, Chau, Chen and Liew showed that the higher
throughput capacity of $\lambda\left(n\right)=\Omega\left(\frac{1}{\sqrt{n}}\right)$
can also be achieved in large-scale CSMA wireless networks if wireless
nodes performing CSMA operations are allowed to use two different
carrier-sensing ranges. The capacity analysis in \cite{Chau11Capacity}
is based on two findings: a) by adjusting the count-down rate, a tunable
parameter in CSMA protocols, of each node, a distributed and randomized
CSMA scheme can achieve the same capacity as a centralized deterministic
scheduling scheme \cite{Jiang10A}; b) by using the highway system
defined in \cite{Franceschetti07Closing}, a higher throughput capacity
of $\lambda\left(n\right)=\Omega\left(\frac{1}{\sqrt{n}}\right)$
can be achieved using a centralized deterministic scheduling algorithm.
Using \cite[Lemma 9]{Chau11Capacity}, which states that in CSMA schemes,
there exists a set of count-down rates such that the throughput of
each and every link is not smaller than that can be achieved with
a centralized deterministic scheduling scheme, and a similar analysis
above for analyzing the capacity of networks in \cite{Franceschetti07Closing},
the result in \cite{Chau11Capacity} can also be obtained using the
relationship established in this paper. Except for some analysis on
particular details of CSMA networks, i.e. hidden node problem and
distributedness of CSMA protocols, the analysis is similar as the
analysis earlier in the section and hence is omitted in the paper.

Observing that in a large network, a much smaller transmission range
is required to connect most nodes in the network (i.e. forming a giant
component) whereas the larger transmission range of $\Theta\left(\sqrt{\frac{\log n}{n}}\right)$
is only required to connect the few hard-to-reach nodes \cite{Ta09On},
a routing scheme can be designed, which achieves a per-node throughput
of $\lambda\left(n\right)=\Theta\left(\frac{1}{\sqrt{n}}\right)$
and does not have to use the highway system, such that a node uses
the smaller transmission ranges for most communications and only uses
the larger transmission if the next-hop node cannot be reached when
using the smaller transmission ranges.

\subsection{Capacity of mobile ad-hoc networks}

In \cite{Grossglauser02Mobility}, Grossglauser and Tse considered
mobile ad hoc networks consisting of $n$ nodes uniformly and i.i.d.
on a unit square $A$ initially. Nodes are mobile and the spatial
distribution of nodes is stationary and ergodic with stationary distribution
uniform on $A$. The trajectories of nodes are i.i.d. Each node chooses
its destination randomly and independently of other nodes. At time
$t$, a node $v_{i}$ can transmit directly to another node $v_{j}$
at rate $W$ if the SINR at $v_{j}$ is above a prescribed threshold
$\beta$:
\[
\frac{P_{i}\left(t\right)\gamma_{ij}\left(t\right)}{N_{0}+\frac{1}{L}\sum_{k\in\Gamma_{i}\left(t\right)}P_{k}\left(t\right)\gamma_{kj}\left(t\right)}>\beta
\]
where $N_{0}$ is the background noise power, $L$ is the processing
gain, $\Gamma_{i}\left(t\right)$ is the set of nodes, not including
$v_{i}$ itself, simultaneously transmitting with $v_{i}$ at time
$t$ and $P_{i}\left(t\right)$ is the transmitting power of $v_{i}$
at time $t$. The transmitting power $P_{i}\left(t\right)$ is determined
by the scheduling algorithm and is chosen to be a constant independent
of $n$. For a narrowband system $L=1$. Parameter $\gamma_{ij}\left(t\right)$
is the channel gain and is given by $\gamma_{ij}\left(t\right)=\left\Vert X_{i}\left(t\right)-X_{j}\left(t\right)\right\Vert ^{-\alpha}$
where $X_{i}\left(t\right)$ represents the location of $v_{i}$ at
time $t$ and $\alpha$ is a parameter greater than $2$.

A two-hop relaying strategy is adopted. In the first phase, a source
transmits a packet to a nearby node (acting as a relay). As the source
moves around, different packets are transmitted to different relay
nodes. In the second phase, either the source or a relay transmits
the packet to the destination when it is close to the destination
and is scheduled to transmit to the destination. Within each time
slot, the set of concurrent transmissions are scheduled randomly and
independently of transmissions in the previous time slot. More specifically,
a parameter $\theta\in\left(0,0.5\right)$, called the transmitter
density, is fixed first. $n_{S}=\theta n$ number of nodes are randomly
designated as transmitters and the remaining nodes are designated
as \emph{potential receivers}. Denote the set of potential receivers
by $R_{t}$. Each transmitter transmits its packets to its nearest
neighbor among all nodes in $R_{t}$. Among all the $n_{S}$ sender-receiver
pairs, only those whose SINR is above $\beta$ are retained. Denote
the number of such pairs by $N_{t}$. Note that the set of transmitter-receiver
pairs is random in each time slot (thus $N_{t}$ is a random integer)
and depends on the time varying locations of nodes. Denote the above
scheduling algorithm by $\chi$.

From the above description of the scheduling algorithm $\chi$, obviously
$1\leq k^{\chi}\left(n\right)\leq2$. Furthermore, it can be shown
\cite[Theorem III-4]{Grossglauser02Mobility} that $Y^{\chi}\left(n\right)=E\left(N_{t}\right)$
and that there exists a positive constant $c$ such that 
\begin{equation}
\lim_{n\rightarrow\infty}\Pr\left(\frac{Y^{\chi}\left(n\right)}{n}\geq c\right)=1\label{eq:MANET lower bound on simultaneous transmission}
\end{equation}
The following result on the asymptotical throughput capacity of the
random mobile ad hoc networks considered in \cite{Grossglauser02Mobility}
readily follows:
\begin{lem}
\label{lem:throughput capacity of mobile ad hoc networks}In the random
mobile ad hoc network considered by Grossglauser and Tse\cite{Grossglauser02Mobility},
a.a.s. $\lambda\left(n\right)=\Theta\left(1\right)$.\end{lem}
\begin{IEEEproof}
We first consider an upper bound of $\lambda\left(n\right)$. It can
be easily shown that $\min_{\chi\in\Phi^{f}}k^{\chi}\left(n\right)=\Omega\left(1\right)$
and $\max_{\chi\in\Phi^{f}}Y^{\chi}\left(n\right)=O\left(n\right)$.
It then follows using Corollary \ref{cor:an upper bound on per-node throughput}
that $\lambda\left(n\right)=O\left(1\right)$.

Now we consider the lower bound. Using the two-phase scheduling algorithm
$\chi$ introduced above, $1\leq k^{\chi}\left(n\right)\leq2$. Using
the above result, (\ref{eq:MANET lower bound on simultaneous transmission})
and Corollary \ref{cor:lower bound on per-node throughput}, conclusion
readily follows that $\lim_{n\rightarrow\infty}\Pr\left(\lambda\left(n\right)\geq\frac{c}{2}W\right)=1$
where $W$ is a constant independent of $n$.
\end{IEEEproof}
The capacity result in \cite{Grossglauser02Mobility} and the use
of the two hop relaying strategy can be intuitively explained as follows.
Obviously the two-hop relaying strategy helps to cap $k^{\chi}\left(n\right)$
at $2$. Compared with a one-hop strategy where a source is only allowed
to transmit when it is close to its destination, the two-hop relaying
strategy also helps to spread the traffic stream between a source-destination
pair to a large number of intermediate relay nodes such that in steady
state, the packets of every source node will be distributed across
all the nodes in the network. This arrangement ensures that every
node in the network will have packets buffered for every other node.
Therefore a node always has a packet to send when a transmission opportunity
is available. Thus the role of the two-hop relaying strategy, compared
with a one-hop strategy is to maximize $Y^{\chi}\left(n\right)$ such
that $Y^{\chi}\left(n\right)=\Theta\left(n\right)$ \cite{Grossglauser02Mobility}
at the expense of a slightly increased $k^{\chi}\left(n\right)$.
A lower bound on $\lambda\left(n\right)$ readily results using $Y^{\chi}\left(n\right)=\Theta\left(n\right)$,
$k^{\chi}\left(n\right)\leq2$ and Corollary \ref{cor:lower bound on per-node throughput}.
An upper bound on $\lambda\left(n\right)$ can be easily obtained
using Corollary \ref{cor:an upper bound on per-node throughput}.
Therefore conclusions readily follows for $\lambda\left(n\right)$.
Capacity of mobile ad-hoc networks assuming other mobility models
and routing strategies \cite{Neely05Capacity} can also be obtained
analogously.

Given the insight revealed in Theorem \ref{thm:Capacity relationship for policy pi}
and Corollaries \ref{cor:capacity of arbitrary networks upper bound},
\ref{cor:capacity of random networks}, \ref{cor:an upper bound on per-node throughput}
and \ref{cor:lower bound on per-node throughput}, it can be readily
shown that in a network with a different traffic model than that in
\cite{Grossglauser02Mobility}, e.g. each node has an infinite stream
of packets for every other node in the network, a one-hop strategy
can also achieve a transport capacity of $\eta\left(n\right)=\Theta\left(n\right)$.
Therefore the insight revealed in Theorem \ref{thm:Capacity relationship for policy pi}
and Corollaries \ref{cor:capacity of arbitrary networks upper bound},
\ref{cor:capacity of random networks}, \ref{cor:an upper bound on per-node throughput}
and \ref{cor:lower bound on per-node throughput} helps to design
the optimum routing strategy for different scenarios of mobile ad-hoc
networks.

\subsection{Multicast capacity}

In the previous three subsections, we have used Theorem \ref{thm:Capacity relationship for policy pi}
and Corollaries \ref{cor:capacity of arbitrary networks upper bound},
\ref{cor:capacity of random networks}, \ref{cor:an upper bound on per-node throughput}
and \ref{cor:lower bound on per-node throughput} established in Section
\ref{sec:Capacity-of-Static-Networks} to analyze the capacity of
the random static and mobile networks considered in \cite{Gupta00the,Franceschetti07Closing,Grossglauser02Mobility}.
An upper bound on the throughput capacity can often be readily obtained
using Corollary \ref{cor:an upper bound on per-node throughput}.
For the lower bound, the procedure generally involves using existing
results and scheduling algorithms already established in \cite{Gupta00the,Franceschetti07Closing,Grossglauser02Mobility}
to obtain $k^{\chi}\left(n\right)$ and $Y^{\chi}\left(n\right)$,
and then using Corollary \ref{cor:lower bound on per-node throughput}
to obtain the throughput capacity lower bound. The use of Theorem
\ref{thm:Capacity relationship for policy pi} and Corollaries \ref{cor:capacity of arbitrary networks upper bound},
\ref{cor:capacity of random networks}, \ref{cor:an upper bound on per-node throughput}
and \ref{cor:lower bound on per-node throughput} often results in
simpler analysis. Similar methods can also be used to obtain the multicast
capacity and capacity of hybrid networks considered in this subsection
and the next subsection. To avoid repetition and to focus on the main
ideas, in this subsection and the next subsection, we choose to give
an intuitive explanation of the results on the multicast capacity
and capacity of hybrid networks only using Theorem \ref{thm:Capacity relationship for policy pi}
and Corollaries \ref{cor:capacity of arbitrary networks upper bound},
\ref{cor:capacity of random networks}, \ref{cor:an upper bound on per-node throughput}
and \ref{cor:lower bound on per-node throughput}.

In \cite{Li09Multicast}, Li considered the multicast capacity of
a network with $n$ nodes uniformly and i.i.d. on a $a\times a$ square,
denoted by $A$. It is assumed that all nodes have the same transmission
range $r\left(n\right)=\Theta\left(\sqrt{\frac{\log n}{n}}\right)$
and are capable of transmitting at $W$ bits per second over a common
channel. Furthermore, a protocol interference model is assumed and
two concurrent transmitters must be separated by an Euclidean distance
of at least $\left(1+\triangle\right)r\left(n\right)$. A subset $S\subseteq V_{n}$
of $n_{s}=\left|S\right|$ nodes are randomly chosen to serve as the
source nodes of $n_{s}$ multicast sessions where $n_{s}$ is assumed
to be sufficiently large. Each node $v_{i}\in S$ chooses a set of
$l-1$ points randomly and independently from $A$ and multicast its
data to the nearest node of each point. Denote by $\Phi^{f}$ the
set of scheduling algorithm that allocate the transport capacity equally
among all multicast sessions. Denote by $\eta^{\chi}\left(n\right)$
the maximum transport capacity that can be achieved \emph{a.a.s.}
using $\chi$. The multicast capacity $\eta\left(n\right)$ is the
maximum transport capacity that can be achieved \emph{a.a.s.} for
all $\chi\in\Phi^{f}$: $\eta\left(n\right)=\max_{\chi\in\Phi^{f}}\eta^{\chi}\left(n\right)$.
Note that a bit multicast to $l-1$ destinations is counted as a single
bit in the calculation of the multicast transport capacity. Therefore
our definition of transport capacity in Section \ref{sec:Network-Models}
is consistent with the definition of the multicast transport capacity
in \cite{Li09Multicast} and the results established in Section \ref{sec:Capacity-of-Static-Networks}
can be used directly here.

We first consider the situation that $l=O\left(\frac{n}{\log n}\right)$.
We will obtain an upper bound on the multicast transport capacity.
It can be readily shown that $\max_{\chi\in\Phi^{f}}Y^{\chi}\left(n\right)=O\left(\frac{1}{r^{2}\left(n\right)}\right)$.
Furthermore, it can be shown that \emph{a.a.s.} any multicast tree
spanning $l$ nodes that are randomly placed in $A$ has a total edge
length of at least $ca\sqrt{l}$ \cite[Lemma 9]{Li09Multicast} where
$c$ is a positive constant. It follows that $\min_{\chi\in\Phi^{f}}k^{\chi}\left(n\right)=\Omega\left(\frac{ca\sqrt{l}}{r\left(n\right)}W\right)$.
Therefore, as an easy consequence of Corollary \ref{cor:an upper bound on per-node throughput},
$\eta\left(n\right)=O\left(\frac{1}{r\left(n\right)\sqrt{l}}W\right)=O\left(\frac{W}{\sqrt{l}}\sqrt{\frac{n}{\log n}}\right)$.

To obtain a lower bound on the multicast transport capacity, a scheduling
algorithm $\chi$ is constructed (see \cite{Li09Multicast} for a
detailed description of the scheduling algorithm $\chi$). More specifically,
$A$ is partitioned into non-overlapping squares and each square is
of size $\frac{r\left(n\right)}{\sqrt{5}}\times\frac{r\left(n\right)}{\sqrt{5}}$.
Calling these squares cells, the total number of cells equals $\frac{5a^{2}}{r^{2}\left(n\right)}$.
Furthermore, nodes located in adjacent cells are directly connected,
where two cells are \emph{adjacent} if they have at least one point
in common. Using the property that nodes are uniformly and i.i.d.,
\emph{a.a.s.} \emph{every} cell has at least one node \cite[Lemma 18]{Li09Multicast}.
Dividing time into time slots of equal length, it can be shown that
there exists a positive integer $c_{1}$, independent of $n$, such
that every $\frac{1}{c_{1}}$ time slots, \emph{every} cell gets at
least one time slot to transit. Using the above results, \emph{a.a.s.}
$Y^{\chi}\left(n\right)\geq\frac{5a^{2}}{c_{1}r^{2}\left(n\right)}$.

Choosing one node from each cell, it can be shown that these nodes
form a connected component, termed \emph{connected dominating set}.
All other nodes are directly connected to at least one node in the
connected dominating set. Multicast traffic is routed using the connected
dominating set. Using the result that for an arbitrary cell, \emph{a.a.s.},
the probability that a randomly chosen multicast flow is routed via
the cell is at most $c_{2}\sqrt{l}r\left(n\right)/a$ \cite[Lemma 20]{Li09Multicast},
\emph{a.a.s. }the number of cells crossed by a randomly chosen multicast
flow is at most $\frac{c_{2}\sqrt{l}r\left(n\right)}{a}\times\frac{5a^{2}}{r^{2}\left(n\right)}=5c_{2}a\frac{\sqrt{l}}{r\left(n\right)}$.
Therefore \emph{a.a.s.} $k^{\chi}\left(n\right)=O\left(\frac{\sqrt{l}}{r\left(n\right)}\right)$
and $\eta^{\chi}\left(n\right)=\Omega\left(\frac{1}{r\left(n\right)\sqrt{l}}W\right)=\Omega\left(\frac{W}{\sqrt{l}}\sqrt{\frac{n}{\log n}}\right)$.

Combing the upper and lower bounds on the transport capacity, conclusion
can be obtained that when $l=O\left(\frac{n}{\log n}\right)$, \emph{a.a.s.}
$\eta\left(n\right)=\Theta\left(\frac{W}{\sqrt{l}}\sqrt{\frac{n}{\log n}}\right)$.

When $l=\Omega\left(\frac{n}{\log n}\right)$, the situation becomes
slightly different. More specifically, the density of the multicast
destination nodes becomes high enough such that the probability that
a single transmission will deliver the packet to more than one multicast
destination nodes becomes high. In fact, using the above connected
dominating set, it can be shown that \emph{a.a.s. }the number of transmissions
required to deliver a packet to all nodes (hence the $l-1$ multicast
destination nodes) is at most $\frac{5a^{2}}{r^{2}\left(n\right)}$,
which is independent of $l$. Consequently $k\left(n\right)=\Theta\left(\frac{1}{r^{2}\left(n\right)}\right)$.
Conclusion then readily follows that when $l=\Omega\left(\frac{n}{\log n}\right)$,
$\eta\left(n\right)=\Theta\left(W\right)$.

\subsection{Capacity of hybrid networks\label{sub:Capacity-of-Hybrid}}

Now we consider the impact of infrastructure nodes on network capacity.
In addition to $n$ ordinary nodes uniformly and i.i.d. on a unit
square $A$, a set of $M$ infrastructure nodes are regularly or randomly
placed in the same area $A$ where $M\leq n$. These infrastructure
nodes act as relay nodes only and do not generate their own traffic.
Following the same setting as that in \cite{Zemlianov05Capacity},
it is assumed that the infrastructure nodes have the same transmission
range $r\left(n\right)=\Theta\left(\sqrt{\frac{\log n}{n}}\right)$
and link capacity $W$ when they communicate with the ordinary nodes
and these infrastructure nodes are inter-connected via a backbone
network with much higher capacity. Furthermore a protocol interference
model is adopted. 

The routing algorithm used in the above network \cite{Zemlianov05Capacity}
has been optimized such that these infrastructure nodes do not become
the bottleneck, which may be possibly caused by a poorly designed
routing algorithm diverting excessive amount of traffic to the infrastructure
nodes.

First consider the case when $M=o\left(\frac{1}{r^{2}\left(n\right)}\right)=o\left(\frac{n}{\log n}\right)$.
In this situation, the number of transmissions involving an infrastructure
node as a transmitter or receiver is small and has little impact on
$Y\left(n\right)$, which has been shown in previous subsections to
be $\Theta\left(\frac{1}{r^{2}\left(n\right)}\right)$. Furthermore,
it can be shown that the average Euclidean distance between a randomly
chosen pair of infrastructure nodes is $\Theta\left(1\right)$ \cite{Philip07The}.
That is, a packet transmitted between two infrastructure nodes moves
by an Euclidean distance of $\Theta\left(1\right)$ whereas a packet
transmitted by a pair of directly connected ordinary nodes moves by
an Euclidean distance of $\Theta\left(r\left(n\right)\right)$. Therefore
a transmission between two infrastructure nodes \emph{is equivalent
to} $\Theta\left(\frac{1}{r\left(n\right)}\right)$ transmissions
between ordinary nodes and the \emph{equivalent }average number of
simultaneous ordinary node transmissions equals $\Theta\left(\left(\frac{1}{r^{2}\left(n\right)}-M\right)+\frac{M}{r\left(n\right)}\right)=\Theta\left(\frac{1}{r^{2}\left(n\right)}+\frac{M}{r\left(n\right)}\right)$.
It follows using a similar procedure outlined in Section \ref{sub:Capacity-of-static-Kumar}
that
\[
\eta\left(n\right)=\Theta\left(\frac{\left(\frac{1}{r^{2}\left(n\right)}+\frac{M}{r\left(n\right)}\right)W}{\frac{1}{r\left(n\right)}}\right)=\Theta\left(\left(\sqrt{\frac{n}{\log n}}+M\right)W\right)
\]
Therefore when $M=o\left(\sqrt{\frac{n}{\log n}}\right)$, the infrastructure
nodes have little impact on the order of $\eta\left(n\right)$; when
$M=\Omega\left(\sqrt{\frac{n}{\log n}}\right)$ (and $M=o\left(\frac{n}{\log n}\right)$),
the infrastructure nodes start to have dominant impact on the network
capacity and the above equation on the transport capacity reduces
to $\eta\left(n\right)=\Theta\left(MW\right)$. Noting that the fundamental
reason why infrastructure nodes improve capacity is that they help
a pair of ordinary nodes separated by a large Euclidean distance to
leapfrog some very long hops, thereby reducing $k\left(n\right)$.
Therefore the same result in the above equation can also be obtained
by analyzing the reduction in $k\left(n\right)$ directly. The analysis
is albeit more complicated.

When $M=\Omega\left(\frac{n}{\log n}\right)$, assuming that the transmission
range stays the same as when $M=o\left(\frac{n}{\log n}\right)$ at
$r\left(n\right)=\Theta\left(\sqrt{\frac{\log n}{n}}\right)$, the
number of simultaneous active infrastructure nodes becomes limited
by the transmission range. More specifically, only $\Theta\left(\frac{1}{r^{2}\left(n\right)}\right)=\Theta\left(\frac{n}{\log n}\right)$
infrastructure nodes can be active simultaneously. Furthermore, \emph{a.a.s.}
each ordinary node can access its nearest infrastructure node in $\Theta\left(1\right)$
hops. Following a similar analysis as that in the last paragraph,
it can be shown that $\eta\left(n\right)=\Theta\left(\frac{nW}{\log n}\right)$
when $M=\Omega\left(\frac{n}{\log n}\right)$.

The above results are consistent with the results in \cite{Zemlianov05Capacity}. 

However we further note that when $M=\Omega\left(\frac{n}{\log n}\right),$
a smaller transmission range of $r\left(n\right)=\Theta\left(\frac{1}{\sqrt{M}}\right)$
is sufficient for an ordinary node to reach its nearest infrastructure
node and hence achieving connectivity. A smaller transmission range
helps to increase $Y\left(n\right)$ and it has been shown previously
that $Y\left(n\right)=\Theta\left(\frac{1}{r^{2}\left(n\right)}\right)$,
while $k\left(n\right)=\Theta\left(1\right)$. Therefore the achievable
transport capacity using the smaller transmission range is $\eta\left(n\right)=\Theta\left(MW\right)=\Omega\left(\frac{nW}{\log n}\right)$,
which is better than the result $\eta\left(n\right)=\Theta\left(\frac{nW}{\log n}\right)$
in \cite{Zemlianov05Capacity}. Moreover, different from the conclusion
in \cite{Zemlianov05Capacity} suggesting that when $M=\Omega\left(\frac{n}{\log n}\right)$,
further investment in infrastructure nodes will not lead to improvement
in capacity, our result suggests that even when $M=\Omega\left(\frac{n}{\log n}\right),$
capacity still keeps increasing linearly with $M$. This capacity
improvement is achieved by reducing the transmission range with the
increase in $M$.

\section{Conclusion and Further Work \label{sec:Conclusion-and-Further}}

In this paper, we show that the network capacity can be determined
by estimating the three parameters, viz. the average number of simultaneous
transmissions, the link capacity and the average number of transmissions
required to deliver a packet to its destination. Our result is valid
for both finite networks and asymptotically infinite networks. We
have demonstrated the usage and the applicability of our result by
using the result to analyze the capacity of a number of different
networks studied in the literature. The use of our result often simplifies
analysis. More importantly, we showed that the same methodology can
be used to analyze the capacity of networks under different conditions.
Therefore our work makes important contributions towards developing
a generic methodology for network capacity analysis that is applicable
to a variety of different scenarios. Furthermore, as illustrated in
Section \ref{sub:Capacity-of-Hybrid}, the simple capacity-determing
relationship revealed in the paper can be used as a powerful and convenient
tool to quickly estimate the capacity of networks based on an intuitive
understanding of the networks. However we readily acknowledge that
the analysis of the three parameters: the average number of simultaneous
transmissions, the link capacity and the average number of transmissions
required to deliver a packet to its destination, may still need some
customized analysis that takes into account details of a network different
from other networks.

For asymptotically infinite random networks, the use of our result
to estimate the capacity often involves estimating the capacity upper
bound and the capacity lower bound separately. The capacity upper
bound can be readily obtained by estimating the maximum number of
simultaneously active transmissions satisfying the interference constraints
that can be accommodated in the network area and the minimum number
of transmissions required to deliver a packet. The capacity lower
bound is more difficult to find. It usually involves constructing
a spatial and temporal scheduling algorithm for the particular network
and demonstrating that the network capacity is achievable using that
algorithm. It remains to be investigated on whether a generic technique
can be found such that the capacity lower bound can be obtained without
resorting to designing customized algorithm for a particular network.

In this paper, we have ignored physical layer details by assuming
that each node is capable of transmitting at a fixed and identical
data rate. This assumption allows us to focus on the topological aspects
of networks that determine capacity. It remains to be investigated
on how to develop a generic methodology to incorporate the impact
of physical layer techniques, e.g. coding and MIMO, on capacity. We
refer readers to recent work by Jiang et al. \cite{Jiang12Towards},
which suggests a possible direction to extend our result to incorporate
physical layer details.

\bibliographystyle{ieeetr}

\begin{IEEEbiography}
[{\includegraphics[width=1in,height=1.25in,clip,keepaspectratio]{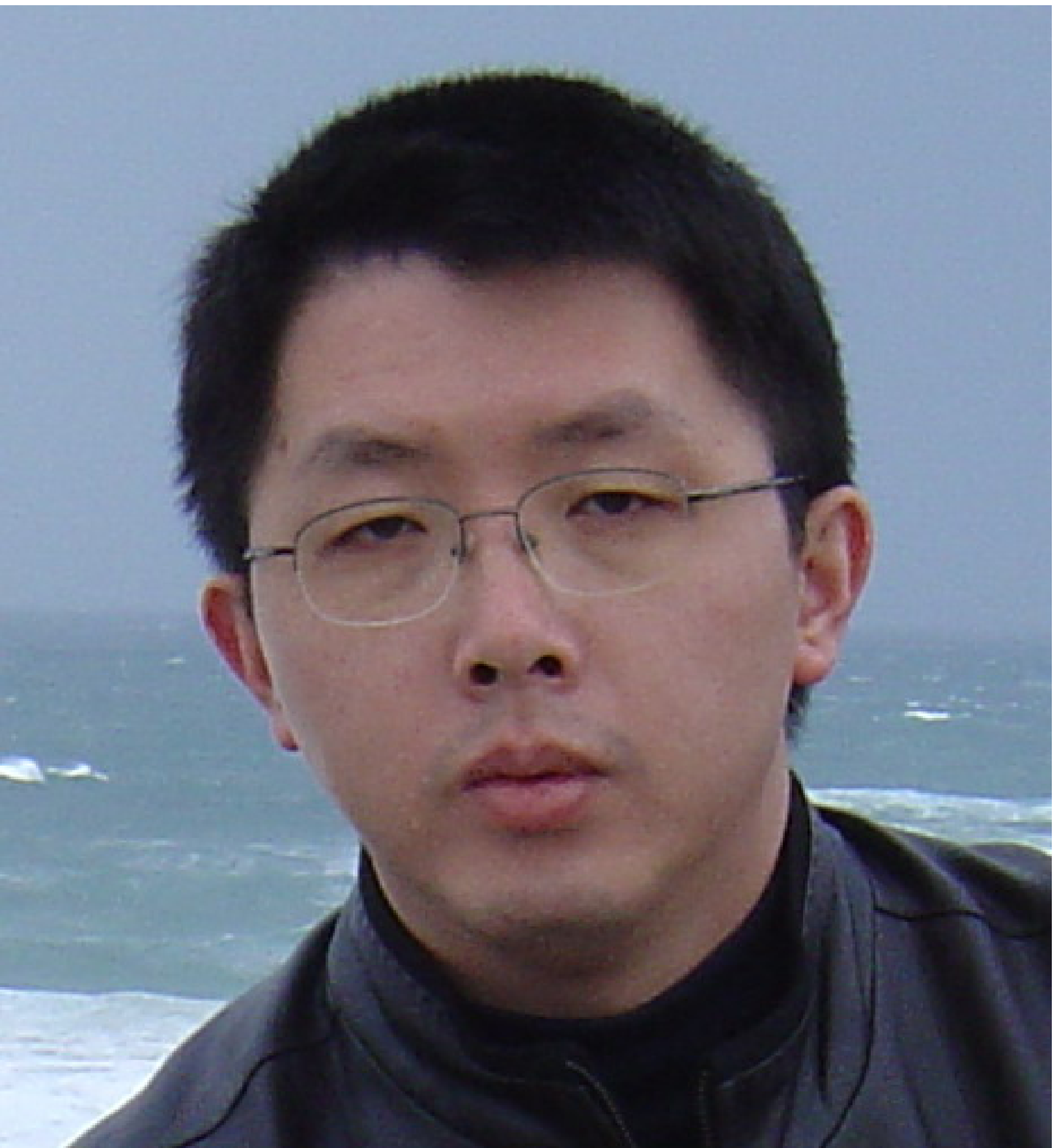}}]{Guoqiang~Mao}
received PhD in telecommunications engineering in 2002 from Edith Cowan University. He joined the School of Electrical and Information Engineering, the University of Sydney in December 2002 where he is an Associate Professor now. His research interest includes intelligent transport systems, applied graph theory and its applications in networking, wireless multihop networks, wireless localization techniques and network performance analysis. He is a Senior Member of IEEE and an Editor of IEEE Transactions on Vehicular Technology.
\end{IEEEbiography}

\begin{IEEEbiography}
[{\includegraphics[width=1in,height=1.25in,clip,keepaspectratio]{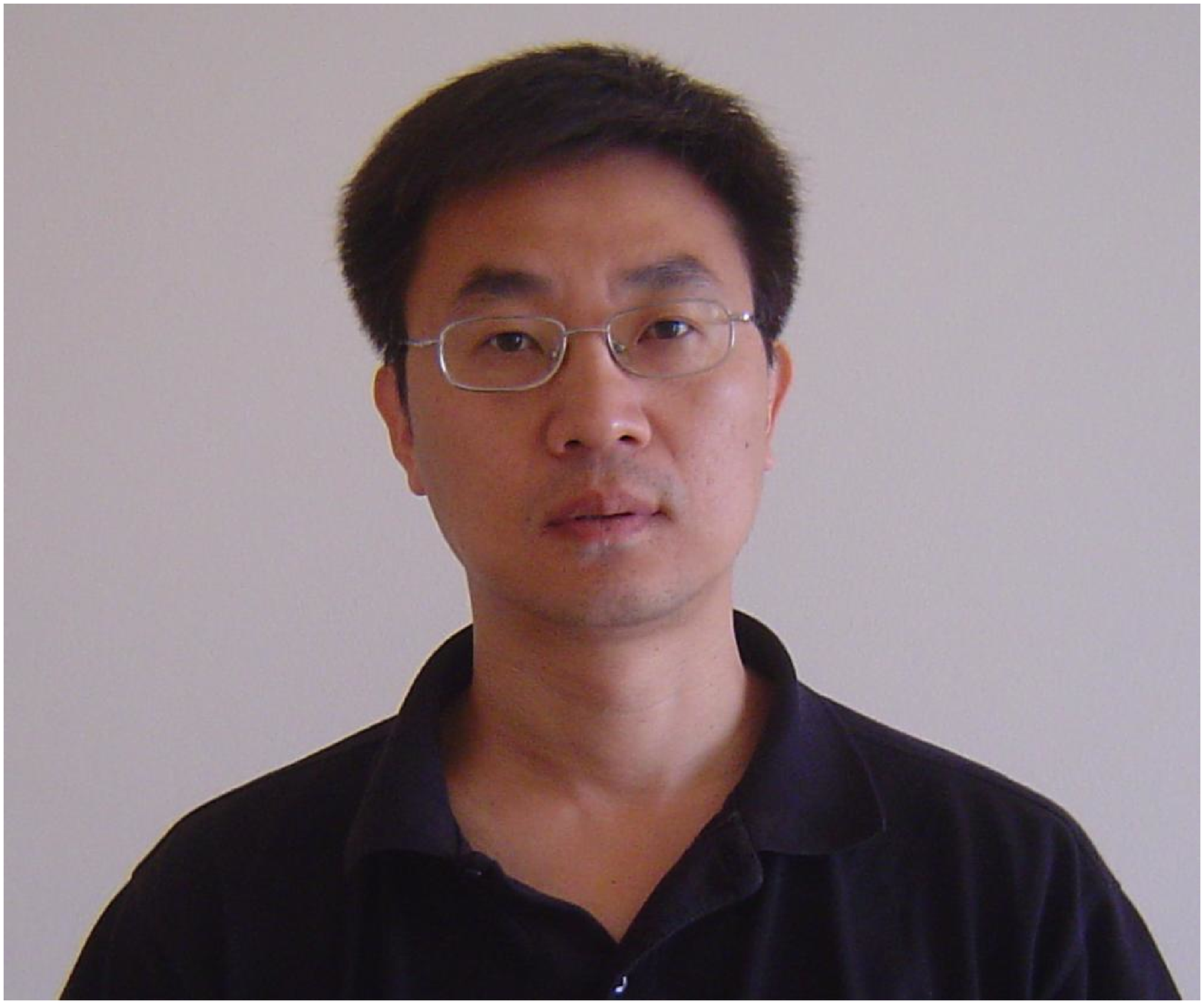}}]{Zihuai~Lin}  received the Ph.D. degree in Electrical Engineering from Chalmers University of Technology, Sweden, in 2006. Prior to this he has held positions at Ericsson Research, Stockholm, Sweden. Following Ph.D. graduation, he worked as a Research Associate Professor at Aalborg University, Denmark and currently as a Research Fellow at the School of Electrical and Information Engineering, the University of Sydney, Australia. His research interests include source/channel/network coding, coded modulation, MIMO, OFDMA, SC-FDMA, radio resource management, cooperative communications, HetNets etc.
\end{IEEEbiography}

\begin{IEEEbiography}
[{\includegraphics[width=1in,height=1.25in,clip,keepaspectratio]{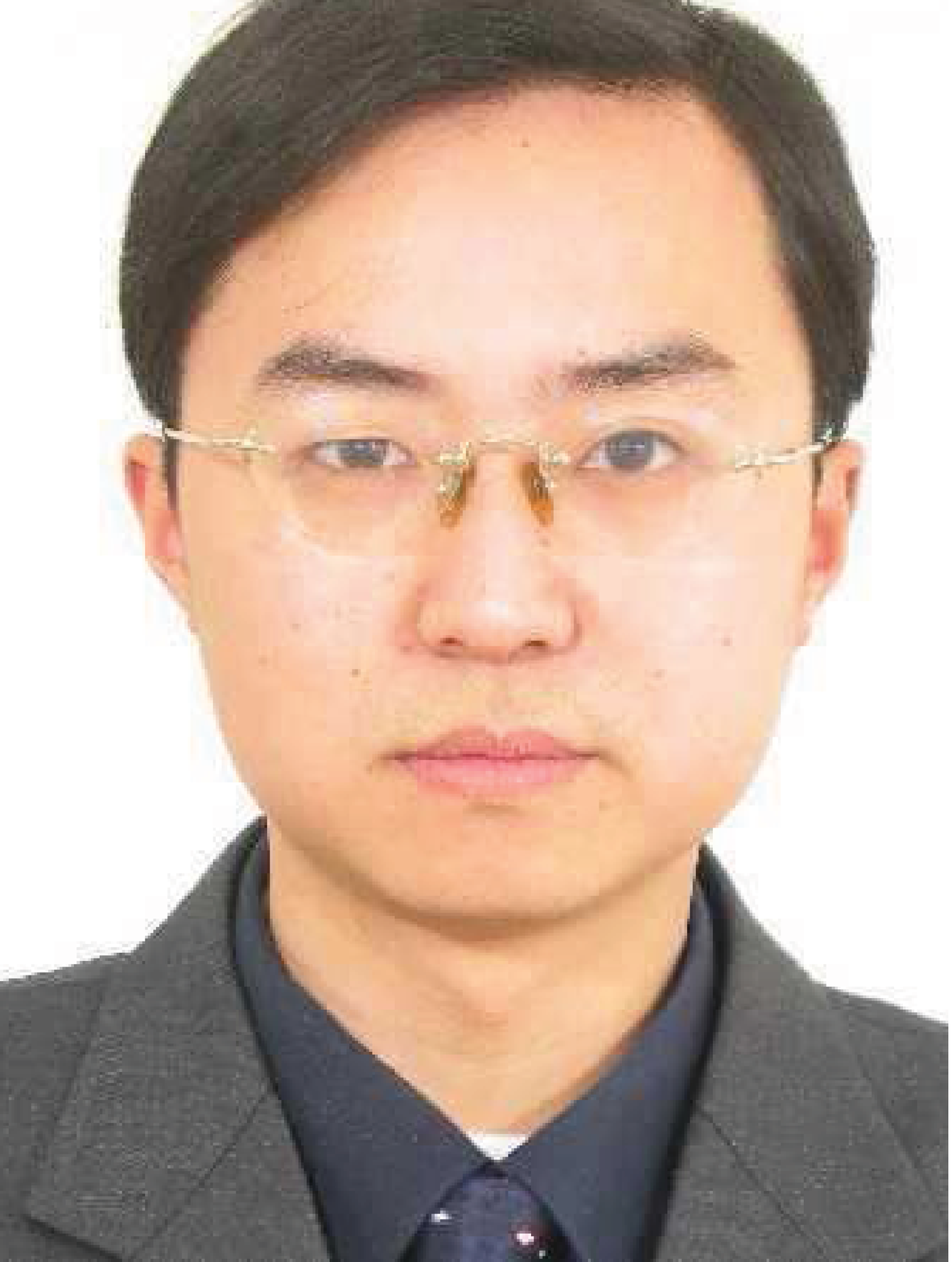}}]{Xiaohu~Ge}
(M'09-SM'11) is currently a Professor with the Department of Electronics and Information Engineering at Huazhong University of Science and Technology (HUST), China. He received his PhD degree in Communication and Information Engineering from HUST in 2003. He has worked at HUST since Nov. 2005. Prior to that, he worked as a researcher at Ajou University (Korea) and Politecnico Di Torino (Italy) from Jan. 2004 to Oct. 2005. He was a visiting researcher at Heriot-Watt University, Edinburgh, UK from June to August 2010. His research interests are in the area of mobile communications, traffic modeling in wireless networks, green communications, and interference modeling in wireless communications. He has published about 60 papers in refereed journals and conference proceedings and has been granted about 15 patents in China. He received the Best Paper Awards from IEEE Globecom 2010. He is leading several projects funded by NSFC, China MOST, and industries. He is taking part in several international joint projects, such as the RCUK funded UK-China Science Bridges: R\&D on (B)4G Wireless Mobile Communications and the EU FP7 funded project: Security, Services, Networking and Performance of Next Generation IP-based Multimedia Wireless Networks.

Dr. Ge is currently serving as an Associate Editor for \textit{International Journal of Communication Systems (John Wiley \& Sons)}, \textit{IET Networks}, \textit{IET Wireless Sensor Systems}, \textit{KSII Transactions on Internet and Information Systems} and \textit{Journal of Internet Technology}. Since 2005, he has been actively involved in the organisation of more than 10 international conferences, such as Executive Chair of IEEE GreenCom 2013 and Co-Chair of workshop of Green Communication of Cellular Networks at IEEE GreenCom 2010. He is a Senior Member of the IEEE, a Senior member of the Chinese Institute of Electronics, a Senior member of the China Institute of Communications, and a member of the NSFC and China MOST Peer Review College.
\end{IEEEbiography}

\begin{IEEEbiography}
[{\includegraphics[width=1in,height=1.25in,clip,keepaspectratio]{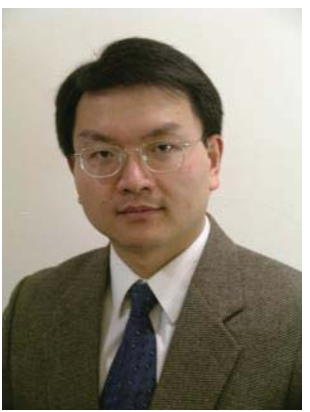}}]{Yang~Yang} received the BEng and MEng degrees in Radio Engineering from Southeast University, Nanjing, P. R. China, in 1996 and 1999, respectively; and the PhD degree in Information Engineering from The Chinese University of Hong Kong in 2002.

Dr. Yang Yang is currently a professor at Shanghai Institute of Microsystem and Information Technology (SIMIT), Chinese Academy of Sciences, and serving as the Director at Shanghai Research Center for Wireless Communications (WiCO). He is a also member of the Founding Team for the School of Information Science and Technology at ShanghaiTech University, which is jointly established by the Shanghai Municipal Government and the Chinese Academy of Sciences (CAS). Prior to that, he served the Department of Electronic and Electrical Engineering at University College London (UCL), United Kingdom, as a Senior Lecturer, the Department of Electronic and Computer Engineering at Brunel University, United Kingdom, as a Lecturer, and the Department of Information Engineering at The Chinese University of Hong Kong as an Assistant Professor. His general research interests include wireless ad hoc and sensor networks, wireless mesh networks, next generation mobile cellular systems, intelligent transport systems, and wireless testbed development and practical experiments.
\end{IEEEbiography}

\end{document}